\documentclass[aps,prl,twocolumn,superscriptaddress,preprintnumbers,amsmath,amssymb,nofootinbib]{revtex4-2}
\usepackage{comment} 
\usepackage{cancel}
\usepackage{graphicx}
\usepackage{dcolumn}
\usepackage{bm}
\usepackage[compact]{titlesec}
\usepackage{float}

\makeatletter
\DeclareRobustCommand{\cev}[1]{%
  {\mathpalette\do@cev{#1}}%
}
\newcommand{\do@cev}[2]{%
  \vbox{\offinterlineskip
    \sbox\z@{$\m@th#1 x$}%
    \ialign{##\cr
      \hidewidth\reflectbox{$\m@th#1\vec{}\mkern4mu$}\hidewidth\cr
      \noalign{\kern-\ht\z@}
      $\m@th#1#2$\cr
    }%
  }%
}
\makeatother
\usepackage{xcolor}

\definecolor{darkbrown}{RGB}{100, 20, 10}

\usepackage[colorlinks=true,linkcolor=blue,citecolor=blue,urlcolor=blue]{hyperref}
\renewcommand{\arraystretch}{2.3}       
\newcommand{\half}{{{\textstyle\frac{1}{2}}}}
\newcommand{\quarter}{{{\textstyle\frac{1}{4}}}}
\newcommand{\be}{\begin{equation}}
\newcommand{\ee}{\end{equation} }
\newcommand{\beqa}{\begin{eqnarray} }
\newcommand{\eeqa}{\end{eqnarray} }
\newcommand{\ba}{\begin{array}}
\newcommand{\ea}{\end{array}}
\newcommand{\bpm}{\begin{pmatrix}}
\newcommand{\epm}{\end{pmatrix}}

\newcommand{\Orth}{\mathbf{O}}

\newcommand{\SU}{\mathbf{SU}}
\newcommand{\Spin}{\mathbf{Spin}}

\newcommand{\rmd}{{\rm d}}
\newcommand{\rd}{{\rmd}}

\newcommand\hcL{{\hat{\cal L}}}

\newcommand\fcL{{{\widetilde{\cal L}}}}

\newcommand{\ODD}{\Orth(D,D)}

\newcommand\So{S_{\scriptscriptstyle{{(0)}}}}
\newcommand\brV{{\bar{V}}}
\newcommand\To{T_{\scriptscriptstyle{{(0)}}}}
\newcommand\KCFT{K^{\rm CFT}}
\newcommand\TCFT{T^{\rm CFT}_{\scriptscriptstyle(0)}}

\newcommand{\SL}{\mathbf{SL}}

\newcommand\Tr{{\scalebox{0.9}{${\mathrm{Tr}}$}}}

\newcommand\cA{{\cal A}}
\newcommand\cB{{\cal B}}

\newcommand\cD{{\cal D}}

\newcommand\cF{{\cal F}}

\newcommand\cH{{\cal H}}
\newcommand\cI{{\cal I}}
\newcommand\cJ{{\cal J}}

\newcommand\cL{{\cal L}}

\newcommand\cN{{\cal N}}
\newcommand\cO{{\cal O}}

\newcommand\cW{{\cal W}}

\def\na{\nabla}

\def\bra{\bar{a}}
\def\brb{\bar{b}}

\def\brbeta{\bar{\beta}}

\def\brp{{\bar{p}}}
\def\brq{{\bar{q}}}

\def\brP{\bar{P}}

\def\brV{\bar{V}}

\newcommand{\trd}{{\bigtriangledown}}

\begin{document}


\title{Non-Riemannian Holography: Long Strings and Soft Hair}

\author{Shaun D. Hampton}
\affiliation{Korea Institute for Advanced Study, 85 Hoegi-ro, Dongdaemun-gu, Seoul 02455, Korea}
\author{Hyun-Cheol Kim}
\affiliation{Department of Physics, Sogang University, 35 Baekbeom-ro, Mapo-gu, Seoul 04107, Korea}
\author{Jae-Hyuk Oh}
\affiliation{Department of Physics, Hanyang University, 222 Wangsimni-ro, Seongdong-gu, Seoul 04763, Korea}
\author{Jeong-Hyuck Park}
\affiliation{Department of Physics, Sogang University, 35 Baekbeom-ro, Mapo-gu, Seoul 04107, Korea}

\begin{abstract}
\vspace{0.3em}
\centering
 \makebox[\linewidth][c]{\begin{minipage}{\dimexpr\paperwidth-4.90cm\relax}
 \looseness=-1\relax\setlength{\parfillskip}{0pt plus 1fil}
We propose an $\mathbf{O}(D,D)$-covariant holographic dictionary
for two bulk completions of the Gomis--Ooguri boundary data:
Riemannian $AdS_3$ supported by pure NS--NS flux and an everywhere
non-Riemannian string Newton--Cartan geometry. An $\mathbf{O}(D,D)$-covariant Codazzi identity
yields the boundary conservation law. Both branches share
diagonal generalized-metric one-point functions, but only the non-Riemannian branch
carries independent normalizable soft hair that couples to long
strings without contributing to surface charges or the renormalized on-shell
action. Supersymmetric non-Riemannian saddles admit arbitrary
chiral Killing-spinor profiles, realizing ``supersupersymmetry.''
\end{minipage}}
\end{abstract}

\maketitle

\setcounter{secnumdepth}{1}
\renewcommand{\thesection}{\arabic{section}}
\titleformat{\section}[block]
 {\centering\normalsize\bfseries}{\thesection.}{0.5em}{}
\section{Introduction}
String theory on $AdS_{3}$ supported by pure NS--NS flux contains macroscopic
closed strings that wind near the boundary: \emph{long
strings}~\cite{Giveon:1998ns,Maldacena:2000hw}.  The near-boundary
Kalb--Ramond coupling cancels the leading energy contribution from string tension,
allowing arbitrarily large strings at finite energy.  Their
spectrum is described by spectrally flowed $\SL(2,\mathbb R)$ WZW
representations~\cite{Giveon:1998ns,Maldacena:2000hw,Maldacena:2000part2}.
At supersymmetric flux level ${k=1}$, long strings become tensionless~\cite{Giribet:2018ada},
and the worldsheet spectrum matches that of the symmetric-product
spacetime CFT~\cite{Eberhardt:2018ouy,Eberhardt:2019ymp}.  Near the boundary,
their dynamics admits a Wakimoto first-order description~\cite{Giveon:1998ns,Knighton:2024,Knighton:2024b},
whose free-field limit is the Gomis--Ooguri
system~\cite{Gomis:2000bd,Danielsson:2000gi}.  Ref.~\cite{Danielsson:2000mu}
recovered the Gomis--Ooguri description of long strings near the boundary
of the Riemannian supergravity dual of noncommutative open-string theory.
Here we seek bulk completions of the NS--NS $AdS_3$ long-string sector,
including an everywhere non-Riemannian one.
Conventional supergravity contains no elementary winding-string degrees of
freedom, while the residual Gomis--Ooguri worldsheet theory has no locally
invertible Riemannian target-space metric.  What bulk
geometry does a long string see?

Double Field Theory (DFT) provides a natural language for this question.
Holographic bulk dynamics is conventionally formulated in
supergravity~\cite{Maldacena:1997re,Gubser:1998bc,Witten:1998qj}, treating
$g_{\mu\nu}$ as geometry and $B_{\mu\nu},\phi$ as matter,
even though T-duality mixes $\{g,B,\phi\}$.
DFT instead treats the massless NS--NS sector
geometrically~\cite{Siegel:1993th,Siegel:1993xq,Hull:2009mi,Hohm:2010pp,Jeon:2010rw,Jeon:2011cn}:
$g$ and $B$ form a constrained generalized metric $\cH_{MN}$ obeying
${\cH\cJ^{-1}\cH=\cJ}$, with $\cJ_{MN}$ the $\Orth(D,D)$-invariant metric,
while $\phi$ and the metric determinant define the $\Orth(D,D)$-singlet
DFT dilaton $d$. Here $D$ denotes the spacetime dimension.
The Einstein double field equation (EDFE) ${G_{MN}=T_{MN}}$, with the gravitational
coupling absorbed into ${T_{MN}}$, unifies the metric, Kalb--Ramond, and
dilaton equations. The DFT Einstein tensor $G_{MN}$, first introduced in
Ref.~\cite{Park:2015noether}, is built covariantly from the NS--NS
gravitational fields $\cH_{MN}$ and $d$ and conserved off-shell,
while $T_{MN}$ is the stringy energy-momentum tensor of other sectors,
conserved on-shell~\cite{Angus:2018mep}.
We follow the conventions and core DFT formalism of Ref.~\cite{Park:2025core}.

The constraint on $\cH_{MN}$ also admits non-Riemannian
parametrizations~\cite{Morand:2017fnv}.  Such backgrounds have no invertible
spacetime metric, even locally, yet strings
propagate consistently with chiral or antichiral constraints along the
degenerate directions~\cite{Lee:2013hma,Ko:2015rha,Morand:2017fnv}.
Constant non-Riemannian vacua are anomaly-free in the critical dimension,
and their BRST spectrum reproduces the linearized EDFE~\cite{Park:2020ixf}.  Singular
nonrelativistic limits of the NS--NS sector lead to string Newton--Cartan
geometry~\cite{Bergshoeff:2019pij,Harmark:2019relating,Bergshoeff:2021nsns,Bidussi:2021ujm},
whereas DFT encodes the same data in a finite non-Riemannian generalized
metric~\cite{Morand:2017fnv,Cho:2019ofr,Gallegos:2020dft,Lescano:2024nr}.
The constant non-Riemannian background of type $(1,1)$, with one chiral
and one antichiral direction, is the target geometry of the
Gomis--Ooguri string~\cite{Ko:2015rha,Morand:2017fnv}.
Such backgrounds can admit ``supersupersymmetry''~\cite{Blair:2020gnh};
see also a nonconstant example in
eleven-dimensional nonrelativistic supergravity~\cite{Blair2025}.

Here we construct Riemannian (R) and non-Riemannian (NR) bulk
completions of the same Gomis--Ooguri boundary geometry and propose
an $\ODD$-covariant holographic dictionary. We derive an
$\ODD$-covariant completion of the Codazzi identity,
which yields the boundary conservation law on-shell, and obtain
the boundary source--response relations and asymptotic symmetry
algebras. We also determine the exact supersymmetries of both branches, including
supersupersymmetry in nonconstant NR bulk backgrounds.

\section{Riemannian Saddle}
\label{sec:R}
We first recast known $AdS_{3}$
holography~\cite{Brown:1986nw,Gubser:1998bc,Witten:1998qj} in the DFT
variables $\{\cH_{MN},d\}$ as a reference for the non-Riemannian dictionary.
The $\{g,B,\phi\}$ decomposition below serves only to display the
NS--NS Ba\~nados geometries, including the BTZ black
hole~\cite{Banados:1992wn,Banados:1992gq,Banados:1998gg}:
\be
 \begin{aligned}
 \vartheta^\pm&:=e^{y/l}\rd x^\pm-e^{-y/l}L_\mp\rd x^\mp\,,\\
 \rd s_{\rm R}^{\,2}&=\rd y^2-2\vartheta^+\vartheta^-\,,\\
 B_{(2)}&=\left(e^{2y/l}+e^{-2y/l}L_+L_-\right)
             \rd x^-\wedge\rd x^+\,,
 \qquad \phi=\phi_0\,,
 \end{aligned}
 \label{Rfields}
\ee
where $\phi_0$ is constant, $y$ is the radial
coordinate with the conformal boundary at ${y\to\infty}$, and
${L_+=L_+(x^+)}$ and ${L_-=L_-(x^-)}$.
Constant ${L_\pm\geq0}$ describe BTZ geometries: the one-sided cases
with exactly one of $L_\pm$ vanishing are extremal black holes,
while ${L_+=L_-=0}$ gives massless BTZ.
Throughout, $l$ is the $AdS_{3}$ radius, and
${x^{\pm}=(t\pm l\varphi)/\sqrt{2}}$ with ${\varphi\sim\varphi+2\pi}$, so
the boundary line element is $-2\rd x^{+}\rd x^{-}$.  We use the
doubled-coordinate basis
${x^{M}=(\tilde x_{\mu};x^{\mu})
=(\tilde x_+,\tilde x_-,\tilde y;x^+,x^-,y)}$; all doubled tensor
components refer to this fixed basis.  The corresponding DFT
variables are
\be
 \begin{gathered}
 \cH^{\rm R}_{MN}=
\scalebox{0.93}{$ \left(\!\ba{cc}
  g^{-1}&-g^{-1}B\\
  Bg^{-1}&g-Bg^{-1}B\ea\!\right)$}\,,
 \quad\!
 \ba{l}
 e^{-2d_{\rm R}}=e^{-2\phi_0+2y/l}\\[-3mm]
 {}\times\scalebox{0.93}{$
 \left(1-L_+L_-e^{-4y/l}\right)\,,$}\ea
 \end{gathered}
 \label{RDFTfields}
\ee
in the region ${e^{4y/l}>L_{+}L_{-}}$, where
${\sqrt{-g}=e^{2y/l}-L_{+}L_{-}e^{-2y/l}>0}$.  With the
constant $\phi_0$ set to zero, their asymptotic form is
\be
 \cH^{\rm R}_{MN}=\cH^{\infty}_{MN}+\cO(e^{-2y/l})\,,
 \qquad d_{\rm R}=-\frac{y}{l}+\cO(e^{-4y/l})\,,
 \label{Rboundary}
\ee
where the leading asymptotic fields are defined by
\begin{subequations}
\begingroup
\setlength{\arraycolsep}{1pt}
\renewcommand{\arraystretch}{1.0}
\be
 \cH^{\infty}_{MN}
 :=\lim_{y\to\infty}\cH^{\rm R}_{MN}
 =
 \left(\begin{array}{@{}ccc@{\hspace{2pt}}|@{\hspace{2pt}}ccc@{}}
  0&0&0&1&0&0\\
  0&0&0&0&-1&0\\
  0&0&1&0&0&0\\ \hline
  1&0&0&0&0&0\\
  0&-1&0&0&0&0\\
  0&0&0&0&0&1
 \end{array}\right)\,,
 \quad
 d^{\infty}:=-\frac{y}{l}\,.
 \label{boundaryH}
\ee
\endgroup
Restricting~\eqref{boundaryH} to
${x^A=(\tilde x_+,\tilde x_-;x^+,x^-)}$ gives the induced boundary generalized
metric $\cH^{(0)}_{AB}$, which is fully non-Riemannian, of type
$(1,1)$ (string Newton--Cartan)~\cite{Morand:2017fnv}, and is reconstructed from the following
${D=2}$ DFT vielbein representative, defined up to boundary local Lorentz
rotations:
\begingroup
\setlength{\arraycolsep}{2pt}
\renewcommand{\arraystretch}{0.9}
\be
 \scalebox{0.91}{$
 V^{(0)}_{Aa}=\left(\ba{cc}
 \tfrac{1}{\sqrt2}&0\\
 0&0\\
 0&-\sqrt2\\
 0&0
 \ea\right)\,,\qquad
 \brV^{(0)}_{A\bra}=\left(\ba{cc}
 0&0\\
 0&\tfrac{1}{\sqrt2}\\
 0&0\\
 \sqrt2&0
 \ea\right)\,.$}
 \label{Rboundaryframe}
\ee
\endgroup
\end{subequations}
In~\eqref{Rboundaryframe}, the lower local indices are
${a=(\oplus,\ominus)}$ and ${\bra=(\bar\oplus,\bar\ominus)}$.
They are raised with flat metrics whose independent nonzero entries are
${\eta_{\oplus\ominus}=-1}$ and ${\bar\eta_{\bar\oplus\bar\ominus}=+1}$.
The full ${D=3}$ limiting DFT vielbeins appear in the Supplemental Material (SM)~\cite{SM}.

For constant ${L_\pm>0}$, the radial patch ends at the BTZ Killing horizon
${e^{2y/l}=\sqrt{L_+L_-}}$, where $e^{-2d}$ vanishes in these coordinates.

\section{Non-Riemannian Saddle}
\label{sec:NR}

On the same physical section
${\tilde{\partial}^{\mu}=0}$, we retain
${L_\pm=L_\pm(x^\pm)}$ and introduce
\be
 \scalebox{0.86}{${
 (\Pi,q,e^\sigma,\chi)=
 \left(L_+L_-,\,e^{-2y/l}\sqrt{\Pi/2},\,\sqrt{L_+/L_-},
 \,2\sqrt{2}\,\operatorname{arctanh}q\right)\,.
 }$}
 \label{NRvariables}
\ee
For definiteness, we take ${L_\pm>0}$ and ${0<q<1}$, allowing the limits ${L_\pm\to0}$.  Then ${e^\sigma,q,\chi>0}$ and
${e^{\pm\sigma}\sinh(\chi/2)=e^{-2y/l}L_\pm+\cO(e^{-6y/l})}$.
The resulting everywhere non-Riemannian generalized metric is
\begingroup
\renewcommand{\arraystretch}{1.35}
\be
\ba{l}
 \cH_{MN}=\\
 \scalebox{0.866}{$
 \left(\begin{array}{ccc|ccc}
 0&0&0&\cosh\chi&-e^{-\sigma}\sinh\chi&0\\
 0&0&0&e^\sigma\sinh\chi&-\cosh\chi&0\\
 0&0&1&0&0&0\\ \hline
 \cosh\chi&e^\sigma\sinh\chi&0
   &-\cW e^\sigma\sinh\chi&\cW\cosh\chi&0\\
 -e^{-\sigma}\sinh\chi&-\cosh\chi&0
   &\cW\cosh\chi&-\cW e^{-\sigma}\sinh\chi&0\\
 0&0&0&0&0&1
 \end{array}\right)\,.$}
 \ea
 \label{NRHcompact}
\ee
\endgroup
The DFT dilaton is
\be
 \ba{ll}
 e^{-2d}
 =e^{2y/l}(1-q^2)\,,\quad&\quad
 d=-\frac{y}{l}+\ln\cosh\frac{\chi}{2\sqrt2}\,.
 \ea
 \label{NRdilaton}
\ee
The remaining function $\cW(x^+,x^-,y)$ follows from the EDFE.  With ${u_\pm:=L_\pm^{-1/2}}$ and
${\cA_\pm:=\partial_\pm^2u_\pm/u_\pm}$, it is
\be
 \begin{aligned}
 \cW={}&W_0+\frac{W_1\chi}{2\sqrt{\Pi}}
 +\frac{l^2\sqrt{\Pi}}{16}\sum_{s=\pm1}\Big[\,
 \cA_su_s^2\,\cI(\chi)\\
 &-2s\left(\partial_+u_++s\,\partial_-u_-\right)^{2}
 \left(e^{s\chi}-1-s\chi\right)
 \Big]\,,
 \end{aligned}
 \label{NRWgeneral}
\ee
where the radial profile is fixed by
\be
 \cI(\chi)=4\sqrt2\int_0^\chi\!\rd t\,
 \left[\frac{\sinh t}{\sinh(t/\sqrt2)}-\sqrt2\right]\,,
 \label{MainGprofile}
\ee
with ${\cI(0)=\cI'(0)=0}$, while $W_0(x^+,x^-)$ and $W_1(x^+,x^-)$
are arbitrary functions.  Both branches solve
${G_{MN}=2l^{-2}\cJ_{MN}}$.  The square-bracketed terms in~\eqref{NRWgeneral}
vanish for constant $L_\pm$ or in either smooth one-sided limit.
As ${y\to\infty}$,
${\cW=W_0+W_1e^{-2y/l}+\cO(e^{-4y/l})}$.
Thus $W_0$ is a source and $W_1$ a
normalizable mode.  At ${W_0=0}$, this saddle approaches~\eqref{boundaryH} while
remaining of type $(1,1)$ at every radius.  For constant ${L_\pm>0}$, its
finite-radius endpoint is ${q=1}$, where ${e^{-2d}=0}$; this endpoint is distinct from the
Riemannian horizon.

In the string Newton--Cartan description, $L_\pm$ enter the
clock forms, while $\cW$ determines the string
vertex~\cite{Andringa:2012uz,Harmark:2017rpg,Bergshoeff:2018yvt,
Bergshoeff:2019pij,Oling:2022fft,Morand:2017fnv,Cho:2019ofr}.
The SM~\cite{SM} gives the explicit clock forms and Kalb--Ramond two-form representative,
and derives this coupling from the Lorentzian doubled-string action.

\section{(Non-)Riemannian Holography}
\label{sec:holography}
\noindent\textit{Doubled holographic dictionary.}---We use the DFT variational
principle~\cite{Angus:2018mep,Cho:2019ofr,Park:2025core} to propose a
Gubser--Klebanov--Polyakov--Witten (GKPW) source--response dictionary
in $\{\cH,d\}$ for both branches.
Its sources are $\cH^{(0)}$ and $d^{(0)}$, with
${d=-y/l+d^{(0)}+\cdots}$:
\be
\displaystyle
 Z_{\rm DFT}\big[\cH^{(0)}\!,d^{(0)}\big]
 =\!\int_{\raisebox{-0.77ex}{$\scriptstyle\cH^{(0)}\!,\,d^{(0)}$}}
\!\! \!\!\!\!\! \!\!\!\!\! \!\!\!\!\!\cD\cH\,\cD d\,e^{iS[\cH,d]}
 =Z_{\rm CFT}\big[\cH^{(0)}\!,d^{(0)}\big]\,.
 \raisetag{-1.1\baselineskip}
\label{RGKPW}
\ee
At fixed boundary sources, we propose that the two branches are candidate
saddles of a single $Z_{\rm DFT}$ dual to one boundary CFT. Their relative contributions to the bulk path integral
remain to be determined.
Let ${h_{p\brq}:=\delta\cH_{MN}V^{M}{}_{p}\brV^{N}{}_{\brq}}$. The
$\Gamma^2$ form of the DFT action~\cite{Park:2025core} is quadratic in
the torsionless DFT Christoffel symbols $\Gamma_{CAB}$~\cite{Jeon:2011cn}.
For these exact saddles, adding the counterterm
${-(16\pi G)^{-1}(4/l)\int_{y=Y}\rd^2x\,e^{-2d}}$ gives
the renormalized on-shell variation~\cite{SM}
\be
 \resizebox{\columnwidth}{!}{$\displaystyle
 \left.\delta S_{\rm ren}\right|_{\rm EOM}
 =\frac{1}{16\pi G}\int_{y=Y}\!\rd^2x\,e^{-2d}
 \left[h^{p\brq}A^{y}_{p\brq}
 +2\delta d\left(B^y+\frac4l\right)\right]\,.$}
 \label{Rrenvariation}
\ee
Here $A^y_{p\brq}$ and $B^y+4/l$ are the generalized-metric
and scalar radial momenta, respectively, with flat indices on
$\Gamma$ denoting DFT vielbein projections:
\be
 \begin{aligned}
 A^{L}_{p\brq}
 &=V^L{}_p\Gamma^M{}_{M\brq}
   +\brV^L{}_{\brq}\Gamma^M{}_{Mp}
   -\Gamma_p{}^L{}_{\brq}-\Gamma_{\brq}{}^L{}_p\,,\\
 B^L&=4\cH^{LM}\partial_Md-\partial_M\cH^{LM}\,.
 \end{aligned}
 \label{Mainmomenta}
\ee
The Fefferman--Graham gauge fixes ${\cH_{yy}={\cH_{\tilde{y}\tilde{y}}=1}}$
and ${\cH_{yA}=\cH_{\tilde{y}A}=0}$, while leaving
$\cH_{AB}(x^{\pm},y)$ and $d(x^{\pm},y)$ arbitrary.
The momenta satisfy an $\ODD$-covariant completion of the
contracted Codazzi identity~\cite{Papadimitriou:2004ap}, valid off-shell in this gauge:
\be
\na^{A}\big(4V_{[A}{}^{a}\brV_{B]}{}^{\brb}A_{a\brb}^y+\cJ_{AB}B^{y}\big)=2G^{y}{}_{B}\,,
\label{Codazzi}
\ee
where the left-hand side is the intrinsic ${D=2}$ DFT divergence on a
constant-$y$ surface and the right-hand side is twice the
radial--tangential component of the bulk ${D=3}$ DFT Einstein tensor~\cite{Park:2015noether}.
For a saddle with ${d=-y/l+o(1)}$, as
in~\eqref{boundaryH}, the GKPW prescription identifies the finite
boundary one-point functions with the radially rescaled momenta:
\begin{subequations}
\be
 \begin{aligned}
 \langle\KCFT_{a\brb}\rangle
 &=-\frac{1}{32\pi G}\lim_{Y\to\infty}e^{2Y/l}A^{y}_{a\brb}(Y)\,,
 \\
 \langle\TCFT\rangle
 &=\frac{1}{16\pi G}\lim_{Y\to\infty}e^{2Y/l}
 \left[B^y(Y)+\frac4l\right]\,.
 \end{aligned}
\label{RKdef}
\ee
These one-point functions have density weight zero~\cite{Angus:2018mep,Park:2025core}.
The scalar $\TCFT$ is dual to the DFT dilaton and is not
\textit{a priori} the stress-tensor trace.
Let $h^{(0)a\brb}$ denote the leading tangential coefficient of
$h^{p\brq}$. We take functional derivatives with respect to the sources
before setting ${h^{(0)}=d^{(0)}=0}$ on the background~\eqref{boundaryH}.
For general sources, the limits denote finite parts after the required
source-dependent counter\-terms are included~\cite{Skenderis:2002wp}.
On-shell, ${G^y{}_{B}=0}$, so the Codazzi identity~\eqref{Codazzi} and the
dictionary~\eqref{RKdef} yield the conserved DFT energy-momentum
tensor of the boundary CFT,
\be
 \scalebox{0.88}{$\displaystyle
 \langle T^{\rm CFT}_{AB}\rangle:=4V_{[A}{}^{a}\brV_{B]}{}^{\brb}
 \langle\KCFT_{a\brb}\rangle
 -\frac12\cJ_{AB}\langle\TCFT\rangle\,,~\,
 \na^{A}\langle T^{\rm CFT}_{AB}\rangle=0\,.$}
 \label{RDFTconservation}
\ee
\end{subequations}
On the constant boundary background~\eqref{boundaryH}, this conservation
law requires $\langle\KCFT_{\ominus\bar{\oplus}}\rangle$ to be constant,
while leaving $\langle\KCFT_{\oplus\bar{\ominus}}\rangle$ locally
unconstrained. The non-Riemannian hair appears in this latter component.

\noindent\textit{Asymptotic symmetry and charges.}---Asymptotic isometries
combine diffeomorphisms and Kalb--Ramond gauge transformations
that preserve the common boundary geometry.  At ${W_0=0}$, for chiral $\epsilon^\pm(x^\pm)$,
the asymptotic components of ${\xi^M=(\tilde\xi_\mu,\xi^\mu)}$ needed for the charges are
\be
 \begin{gathered}
 \xi^\pm_{\rm R}=\epsilon^\pm
 +\tfrac{l^2}{4}e^{-2y/l}\partial_\mp^2\epsilon^\mp\,,
 \quad \xi^\pm_{\rm NR}=\epsilon^\pm\,,\\
 \tilde\xi_\pm=\mp\tfrac{l^2}{2}e^{-2y/l}
 L_\pm\partial_\mp^2\epsilon^\mp\,,\\
 \xi^y=-\tfrac l2(\partial_+\epsilon^++\partial_-\epsilon^-)\,,
 \quad \tilde\xi_y=-\tfrac l2(\partial_+\epsilon^+-\partial_-\epsilon^-)\,.
 \end{gathered}
 \label{Rkilling}
\ee
The radial and dual components agree at the displayed order.
The Riemannian generator induces
\be
 \delta_{\epsilon}L_{\pm}=\epsilon^{\pm}\partial_{\pm}L_{\pm}
 +2L_{\pm}\partial_{\pm}\epsilon^{\pm}
 -\tfrac{l^{2}}{4}\,\partial_{\pm}^{3}\epsilon^{\pm}\,.
\label{RVirasoro}
\ee
On the NR saddle, the corresponding laws are (with $i$ summed over $\pm$)
\be
 \begin{aligned}
 \delta_{\epsilon}L_{\pm}&=\epsilon^{\pm}\partial_{\pm}L_{\pm}
 +2L_{\pm}\partial_{\pm}\epsilon^{\pm}\,,\\
 \delta_{\epsilon}W_{1}&=\epsilon^{i}\partial_{i}W_{1}
 +2W_{1}\partial_{i}\epsilon^{i}
 -l^{2}\big(L_{-}\partial_{+}^{3}\epsilon^{+}
 +L_{+}\partial_{-}^{3}\epsilon^{-}\big)\,.
 \end{aligned}
 \label{NRasympt}
\ee
The DFT-covariant charges~\cite{Park:2015noether,Blair:2015eba} have the
same expression on both branches,
$Q[\epsilon]=\frac{1}{4\pi Gl}\oint
 (L_+\epsilon^+\rd x^+-L_-\epsilon^-\rd x^-)$.
The restricted classical charge algebras have ${c_{\rm charge}^{\rm R}=3l/(2G)}$
and ${c_{\rm charge}^{\rm NR}=0}$; the latter does not fix an OPE central charge.
$W_1$ is absent from the charges. It also drops out of the
renormalized on-shell action of these exact saddles when no
additional boundary action term is included at the interior endpoint~\cite{SM}.
It is thus charge-neutral soft hair~\cite{Hawking:2016msc,Hawking:2016sgy}.

\noindent\textit{One-point functions.}---The dictionary~\eqref{RKdef}
gives the one-point functions on the two exact saddles, using the chosen
DFT vielbeins~\eqref{Rboundaryframe}:
\begingroup
\renewcommand{\arraystretch}{1.05}
\be
\begin{aligned}
16\pi Gl\,\langle\KCFT_{a\brb}\rangle_{\rm R}
 &=\begin{pmatrix}L_+&L_+L_-\\1&L_-\end{pmatrix}\,,\quad\langle\TCFT\rangle_{\rm R}=0\,,\\
16\pi Gl\,\langle\KCFT_{a\brb}\rangle_{\rm NR}
 &=\begin{pmatrix}L_+&W_1/4\\0&L_-\end{pmatrix}\,,
 \quad\langle\TCFT\rangle_{\rm NR}=0\,.
\end{aligned}
\label{Mainonepoints}
\ee
\endgroup
The rows and columns are ordered as ${a=(\oplus,\ominus)}$ and ${\brb=(\bar\oplus,\bar\ominus)}$, respectively.
Thus, the two saddles share diagonal
generalized-metric one-point functions but differ off the diagonal: only the
non-Riemannian branch carries the independent normalizable hair $W_{1}$.

\noindent\textit{Two-point functions.}---These follow from the second variation of the renormalized on-shell action,
with the sources set to zero only after differentiation.
We order the five insertions as
$\mathcal O=(\KCFT_{\oplus\bar\oplus},\allowbreak\KCFT_{\ominus\bar\ominus},\allowbreak
\KCFT_{\oplus\bar\ominus},\allowbreak\KCFT_{\ominus\bar\oplus},\allowbreak\TCFT)$,
with source components $h^{(0)}_{\ominus\bar\ominus}$,
$h^{(0)}_{\oplus\bar\oplus}$, $h^{(0)}_{\ominus\bar\oplus}$,
$h^{(0)}_{\oplus\bar\ominus}$, and $d^{(0)}$, respectively.
The connected time-ordered correlator matrix is
$(\mathbb G_s)_{IJ}(x,x')
=\langle\mathcal O_I(x)\mathcal O_J(x')\rangle_{s,\mathrm{conn}}$,
where ${s\in\{\mathrm R,\mathrm{NR}\}}$ labels the saddles.

Varying the measure, momenta, and DFT vielbeins
in~\eqref{Rrenvariation} on linearized bulk solutions gives
\begingroup
\renewcommand{\arraystretch}{1.05}
\be
\mathbb G_s=
\begin{pmatrix}
A_+^s&0&(M_+^s)^T&0&0\\
0&A_-^s&(M_-^s)^T&0&0\\
M_+^s&M_-^s&0&0&0\\
0&0&0&0&0\\
0&0&0&0&0
\end{pmatrix}
+\mathcal R_s+\mathcal C_s^{\rm loc}\,.
\label{Mainfivebyfive}
\ee
\endgroup
Transposition means ${F^T(x,x')=F(x',x)}$.
$\mathcal R_s$, including hair--hair correlations, depends on the state
and the interior conditions.
$\mathcal C_s^{\rm loc}$ contains contact terms with finite-counterterm
ambiguities; the displayed zeros do not imply vanishing full correlators.

The vanishing one-point and particular two-point functions
involving $\TCFT$ are consistent with the dilaton decoupling required
to realize the two Witt algebras of $\cH^{(0)}$.
Imposing also ${\hcL_\xi d^{(0)}=0}$ would restrict the boundary
diffeomorphisms to a finite-dimensional subalgebra~\cite{Blair:2020gnh}.
A free supersymmetric boundary CFT realizing these properties is
constructed in the SM~\cite{SM}.

The particular Riemannian mixed kernels obey
${M_\pm^{\rm R}(x,x')=L_\mp(x^\mp)A_\pm^{\rm R}(x,x')}$.
Before imposing the spatial $S^1$ identification and interior
conditions, the linearized EDFE yields the following particular
kernels on the covering space of the boundary cylinder~\cite{SM},
with ${\Delta_\pm=x^\pm-x'^\pm}$:
\be
\resizebox{\columnwidth}{!}{$\displaystyle
\begin{aligned}
A_\pm^s(x,x')&=\frac{1}{128\pi^2Gl}
 \left[\frac{Gl\,c_{\rm charge}^s}{\Delta_\pm^4}
       +\frac{\partial_\pm L_\pm(x^\pm)}{\Delta_\pm}
       -\frac{2L_\pm(x^\pm)}{\Delta_\pm^2}\right]\,,\\
M_\pm^{\rm NR}(x,x')&=\frac{1}{512\pi^2Gl}
 \left[\frac{\partial_\pm W_1(x)}{\Delta_\pm}
       -\frac{2W_1(x)}{\Delta_\pm^2}
       +\frac{6l^2L_\mp(x^\mp)}{\Delta_\pm^4}\right]\,.
\end{aligned}$}
\label{Rcorrelators}
\ee
The independent $W_1$ data thus enter both the off-diagonal one-point
function in~\eqref{Mainonepoints} and the particular mixed kernels
$M_\pm^{\rm NR}$.

At ${L_\pm=0}$, the Riemannian result reproduces
the Poincar\'e-vacuum stress correlator on the covering space: multiplying $A_\pm^{\rm R}$
by $(8\pi)^2$ gives ${c/(2\Delta_\pm^4)}$ and fixes
${c=c_{\rm charge}^{\rm R}=3l/(2G)}$~\cite{Brown:1986nw}.
The particular NR diagonal kernel lacks this fourth-order pole
but retains the terms dictated by the weight-two conformal
transformation of $L_\pm$ in~\eqref{NRasympt}. The $\Delta_\pm^{-4}$ term in $M_\pm^{\rm NR}$ depends on $L_\mp$
and does not define a state-independent central charge.

\noindent\textit{Long strings and hair.}---The worldsheet theory on the Riemannian saddle is the
$\SL(2,\mathbb{R})$ WZW model~\cite{Giveon:1998ns,Maldacena:2000hw}.  Near the boundary,
its long-string description has the Gomis--Ooguri
effective speed of light $c_{\rm eff}$, defined by
${c_{\rm eff}^{\,2}:=
2(e^{2y/l}+L_+L_-e^{-2y/l})\sim2e^{2y/l}}$, so the Gomis--Ooguri limit
${c_{\rm eff}\to\infty}$ is reached without
parameter tuning~\cite{Ko:2015rha,Oling:2022fft,Blair:2020ttbar}.
The two branches share the asymptotic Gomis--Ooguri theory, with
a hair vertex
\be
 \resizebox{\columnwidth}{!}{$\displaystyle
 \mathcal L_{\rm GO}=\frac{\beta\bar\partial x^++\brbeta\partial x^-
 +\partial y\bar\partial y}{2\pi\alpha'}\,,\quad
 V_{\cW}=\frac{\cW(x^+,x^-,y)\partial x^+\bar\partial x^-}{4\pi\alpha'}\,.$}
 \label{Mainworldsheet}
\ee
The Lagrange multipliers enforce ${\bar\partial x^+=0=\partial x^-}$, and
$\cW$ approaches $W_0$ at the boundary. The $W_0$ vertex has conformal weight $(1,1)$ and a nonsingular
self-OPE, making the boundary deformation exactly marginal~\cite{SM}.
The radial zero mode $W_0$ is locally pure gauge, but the
required large gauge transformation acts nontrivially at the boundary
and changes the source.

The other radial mode, $W_1$, is normalizable.
Around the non-Riemannian vacuum, the linearized $W_1$ mode
is not pure gauge.
The variational principle pairs the $W_1$ response
with the type-changing boundary source
${\delta\cH^{(0)+-}=-\half h^{(0)}_{\ominus\bar\oplus}}$.
This source couples to the Gomis--Ooguri bilinear $\beta\bar\beta$,
deforming the type-$(1,1)$ non-Riemannian generalized metric
toward a Riemannian one~\cite{Cho:2019ofr}.  At the common source ${W_0=0}$, the Riemannian branch contains a fixed,
$L_\pm$-independent $e^{-2y/l}$ falloff in $\cH^{+-}$, whereas the
type-$(1,1)$ branch has no such term and may carry $W_1$ hair.
The $W_1$ mode couples directly to the long-string worldsheet through
$V_{\cW}$, yet contributes neither to the asymptotic surface charges
nor to the renormalized on-shell action of these exact saddles under the stated interior prescription.

\noindent\textit{Supersymmetry.}---Adjoining $S^3\times T^4$ with
compensating NS--NS flux gives ${D=10}$ critical uplifts satisfying
${G^{(10)}_{\hat M\hat N}=0}$~\cite{SM}. We use
type-II supersymmetric DFT~\cite{Jeon:2012hp}, with
independent $\mathbf{Spin}(1,9)$ and $\mathbf{Spin}(9,1)$ local Lorentz symmetries.
The Killing-spinor equations are given in the SM~\cite{SM}.
For arbitrary smooth chiral $L_\pm$, the Riemannian uplift
locally admits sixteen independent real Killing-spinor solutions,
eight in each spin sector, before imposing spatial-circle periodicity.
This is the standard local $AdS_3\times S^3$ $1/2$-BPS
count~\cite{Jeon:2012hp,Colgain:2016,Deger:1998}.

The non-Riemannian vacuum~\eqref{boundaryH} with ${L_{\pm}=\cW=0}$ admits
infinite-dimensional supersupersymmetry~\cite{Jeon:2011sq,Hohm:2011nu,Blair:2020gnh},
with arbitrary chiral profiles in sixteen real polarizations and local
closure on supertranslations (cf. Ref.~\cite{Blair2025}). At ${W_0=0}$, the Riemannian massless BTZ saddle and the
non-Riemannian hairy saddle with ${L_\pm=0}$ and ${W_1\neq0}$ preserve complementary sets of
constant, spatially periodic Killing spinors: each saddle
preserves four in the $\mathbf{Spin}(1,9)$ sector and four
in the $\mathbf{Spin}(9,1)$ sector. The two sets have no overlap and together
span all sixteen constant vacuum modes~\cite{Jeon:2012hp,SM}.

The non-Riemannian one-sided branches admit Killing-spinor families with four
arbitrary $x^+$-chiral functions in ${\mathbf{Spin}}(1,9)$ for ${L_-=0}$,
or four arbitrary $x^-$-chiral functions in ${\mathbf{Spin}}(9,1)$ for
${L_+=0}$, independently of the smooth $W_1$ profile~\cite{SM}.
Arbitrary smooth periodic profiles give an infinite-dimensional Killing-spinor
space on the spatial circle; finite counts refer only to polarizations.

\section{Conclusion}

We have introduced an $\ODD$-covariant framework for doubled holography
with Riemannian and non-Riemannian bulk geometries, and proposed
non-Riemannian holography for the Gomis--Ooguri long-string sector.
The DFT analogue of the Codazzi identity~\eqref{Codazzi} underlies
boundary DFT energy-momentum conservation in both branches.
At ${W_0=0}$, the same supersupersymmetric non-Riemannian boundary
geometry~\eqref{boundaryH} supports the Riemannian Ba\~nados family and an everywhere
non-Riemannian branch. They share diagonal
generalized-metric one-point functions, but only the non-Riemannian branch
carries independent normalizable soft hair $W_{1}$, which couples directly to the
long-string worldsheet through $V_{\cW}$ in Eq.~\eqref{Mainworldsheet},
yet carries no covariant surface charge and drops out of the renormalized on-shell
action of these exact saddles under the stated prescription at the interior endpoint.

The Riemannian one-point function in Eq.~\eqref{Mainonepoints} and the
particular mixed-kernel relation ${M_\pm^{\rm R}=L_\mp A_\pm^{\rm R}}$,
with $A_\pm^{\rm R}$ given in Eq.~\eqref{Rcorrelators}, suggest a semiclassical
$T\bar T$-like factorization~\cite{Zamolodchikov:2004ce}.
The relation of this factorization to $T\bar T$ descriptions of nonrelativistic
strings~\cite{Blair:2020ttbar} and to the independent non-Riemannian hair
remains to be clarified.

The two saddles are candidate radial trajectories from the same
supersymmetric fixed point. Their interpretation as distinct Wilsonian RG
flows~\cite{Osborn:1991gm,Erdmenger:2001} remains open. A central
question is whether the non-Riemannian saddle contributes to the bulk
quantum-gravity path integral and, if so, how it modifies results
from a Riemannian-only saddle sum. Whether the normalizable hair $W_{1}$
admits an interpretation as a long-string condensate is also open.
A natural direction is to explore an extension of this framework to
other AdS backgrounds supported by Ramond--Ramond flux within
formulations of supergravity with manifest U-duality~\cite{Berman:2010geometry,Park:2013ugeometry,Hohm:2013pua}.

\noindent{\textit{Acknowledgments.}}---We thank Junho Hong, Seok Kim, and Chanyong Park for helpful discussions. \mbox{OpenAI} ChatGPT/Codex and Anthropic Claude assisted with cross-checks; all AI-assisted output was verified by the authors, who take full responsibility for the content. This work is supported by the National Research Foundation of Korea (NRF) through Grants RS-2023-NR077094 and RS-2020-NR049598 (Center for Quantum Spacetime: CQUeST) and by KIAS Grant PG096301.

\vspace{3pt}
\begingroup
\small
\noindent\href{mailto:sdh2023@kias.re.kr}{\texttt{sdh2023@kias.re.kr}};\enspace~~
\href{mailto:kimhc8888@sogang.ac.kr}{\texttt{kimhc8888@sogang.ac.kr}}\\
\href{mailto:jaehyukoh@hanyang.ac.kr}{\texttt{jaehyukoh@hanyang.ac.kr}};\enspace~~
\href{mailto:park@sogang.ac.kr}{\texttt{park@sogang.ac.kr}}\par
\endgroup


\begingroup
\makeatletter
\renewcommand{\bibsection}{%
  \par
  \begingroup
    \baselineskip8\p@
    \bib@device{\hsize}{72\p@}%
  \endgroup
  \nobreak\@nobreaktrue
  \addvspace{9\p@}%
}
\makeatother

\endgroup

\clearpage

\begin{widetext}
\appendix
\setcounter{equation}{0}
\setcounter{figure}{0}
\setcounter{section}{0}
\setcounter{subsection}{0}
\setcounter{secnumdepth}{2}
\renewcommand{\thesection}{SM\,\arabic{section}}
\renewcommand{\thesubsection}{SM\,\arabic{section}.\arabic{subsection}}
\makeatletter
\@removefromreset{equation}{section}
\@removefromreset{figure}{section}
\renewcommand{\p@subsection}{}
\makeatother
\renewcommand{\theequation}{SM\,\arabic{equation}}
\renewcommand{\thefigure}{SM\,\arabic{figure}}
\titleformat{\section}[block]
 {\centering\Large\bfseries}{\thesection}{0.8em}{}
\titleformat{\subsection}[block]
 {\centering\large\bfseries}{\thesubsection}{0.8em}{}
\titlespacing*{\section}{0pt}{2.5ex plus .5ex}{1.2ex}
\titlespacing*{\subsection}{0pt}{1.8ex plus .4ex}{0.8ex}
\setlength{\jot}{9pt}                 
\renewcommand{\arraystretch}{1.35}

\begin{center}
\vspace{9pt}
{\Large\bfseries Supplemental Material (SM)}\\[8pt]
{\large\bfseries Non-Riemannian Holography: Long Strings and Soft Hair}\\[4pt]
{\normalsize Shaun D. Hampton, Hyun-Cheol Kim, Jae-Hyuk Oh, and Jeong-Hyuck Park}\\[3pt]
{\small\texttt{sdh2023@kias.re.kr;}\enspace
 \texttt{kimhc8888@sogang.ac.kr;}\enspace
 \texttt{jaehyukoh@hanyang.ac.kr;}\enspace
 \texttt{park@sogang.ac.kr}}
\end{center}
\vspace{7pt}
\pagenumbering{roman}

\begin{center}
 {\large\bfseries Contents of the Supplemental Material}
\end{center}
\begingroup
\renewcommand{\arraystretch}{1.25}
\noindent\begin{tabular*}{\textwidth}{@{\extracolsep{\fill}}lr}
 \textbf{\ref{SMsec:action}\quad Bulk Action and Renormalized Saddle Values} & \textbf{\pageref{SMsec:action}}\\
 \quad\ref{SMsub:action:R}\quad Riemannian Saddle Action & \pageref{SMsub:action:R}\\
 \quad\ref{SMsub:action:NR}\quad Non-Riemannian Saddle Action & \pageref{SMsub:action:NR}\\
 \textbf{\ref{SMsec:linear}\quad Linearized Dynamics and Holographic Renormalization} & \textbf{\pageref{SMsec:linear}}\\
 \quad\ref{SMsub:variations}\quad Constrained Variations and Linearized EDFE & \pageref{SMsub:variations}\\
 \quad\ref{SMsub:momenta}\quad Boundary Variations and Counterterms & \pageref{SMsub:momenta}\\
 \quad\ref{SMsub:linear:R}\quad Riemannian Branch: Linearized Solutions & \pageref{SMsub:linear:R}\\
 \quad\ref{SMsub:linear:NR}\quad Non-Riemannian Vielbeins and Linearized Solutions & \pageref{SMsub:linear:NR}\\
 \textbf{\ref{SMsec:worldsheet}\quad Worldsheet Actions and Vertex Operators} & \textbf{\pageref{SMsec:worldsheet}}\\
 \quad\ref{SMsub:worldsheet:R}\quad Riemannian Sigma Model and Long Strings & \pageref{SMsub:worldsheet:R}\\
 \quad\ref{SMsub:worldsheet:NR}\quad Non-Riemannian Worldsheet and Marginal Deformations & \pageref{SMsub:worldsheet:NR}\\
 \textbf{\ref{SMsec:susy}\quad Ten-Dimensional Uplift and Killing Spinors} & \textbf{\pageref{SMsec:susy}}\\
 \quad\ref{SMsub:susy:equations}\quad Ten-Dimensional Killing-Spinor Equations & \pageref{SMsub:susy:equations}\\
 \quad\ref{SMsub:susy:periodic}\quad Killing Spinors under the Periodic Identification & \pageref{SMsub:susy:periodic}\\
 \textbf{\ref{SMsec:candidate}\quad A Free Abelian CFT on the DFT Boundary} & \textbf{\pageref{SMsec:candidate}}\\
\end{tabular*}
\endgroup
\vspace{8pt}

We follow the core conventions of Ref.~\cite{Park:2025core}.
Doubled indices are $M,N,\cdots$ in the three-dimensional bulk,
$A,B,\cdots$ on the two-dimensional boundary, and
$\hat M,\hat N,\cdots$ in the ten-dimensional uplift; their flat
counterparts are $p,q,\cdots$, $a,b,\cdots$, and
$\hat p,\hat q,\cdots$, together with barred indices.
The physical section is ${\tilde\partial^\mu=0}$, with doubled coordinate
order $(\tilde x_+,\tilde x_-,\tilde y;x^+,x^-,y)$.
As in Ref.~\cite{Park:2025core}, $\nabla_M$ denotes the semi-covariant
derivative with the torsionless DFT Christoffel symbols and $\cD_M$ the
master derivative, which includes the two spin connections.

\section{Bulk Action and Renormalized Saddle Values}
\label{SMsec:action}

To evaluate the on-shell action of both saddles, we use the
$\Gamma^{2}$ identity~\eqref{GammaDFT}, with $B^{M}$ as in
Eq.~\eqref{defB}:
\be
 e^{-2d}\So=\cL_{\Gamma^{2}}
 +\partial_{M}\big(e^{-2d}B^{M}\big)\,,
 \qquad
 B^{M}=4\cH^{MN}\partial_{N}d-\partial_{N}\cH^{MN}\,.
 \label{SMgamma2}
\ee
Here $\cL_{\Gamma^{2}}$ is quadratic in the torsionless DFT Christoffel
symbols.  On-shell, ${\So=2\Lambda_{\rm DFT}}$ with ${\Lambda_{\rm DFT}=-2/l^{2}}$, so the
Lagrangian density of the $\Gamma^{2}$ action reduces to a total
derivative,
${\cL_{\Gamma^{2}}-2\Lambda_{\rm DFT}\,e^{-2d}
=-\partial_{M}\big(e^{-2d}B^{M}\big)}$.  On the physical section $\tilde\partial^{\mu}=0$, the
divergence involves only $B^{x^{\pm}}$ and $B^{y}$.  The tangential components are
built only from the $\cW$-independent blocks of $\cH$, and
the associated tangential total derivatives integrate to zero under the periodic or decay conditions used here.
Since ${\cH^{yN}=\delta^N_y}$ on both saddles, for arbitrary $L_{\pm}$
and independently of $\cW$ on the non-Riemannian branch,
\be
 B^{y}=4\,\partial_{y}d\,,
 \qquad
 e^{-2d}B^{y}=-2\,\partial_{y}e^{-2d}
 =-\frac{4}{l}\Big(e^{2y/l}+\mu\,e^{-2y/l}\Big)\,.
 \label{SMgamma2flux}
\ee
Here
\be
 \mu=L_{+}L_{-}\quad({\rm R})\,,\qquad
 \mu=\half L_{+}L_{-}\quad({\rm NR})\,,
 \label{SMmudefinition}
\ee
so the same formula applies to both branches.  The divergent boundary
contribution associated with~\eqref{SMgamma2flux} is canceled by the counterterm
$-(16\pi G)^{-1}(4/l)\int_{y=Y}\rd^2x\,e^{-2d}$; the remaining cutoff
contribution vanishes as the cutoff is removed.  For constant ${L_\pm>0}$ and an interior
endpoint ${e^{2y_*/l}=\sqrt\mu}$, the bulk total derivative integrated up to
the cutoff and the volume counterterm combine as
\be
 \begin{aligned}
 S_{\rm ren}(Y)
 &=\frac{1}{16\pi G}\int\rd^2x\,
 \bigg[\frac{4}{l}\Big(e^{2Y/l}+\mu e^{-2Y/l}\Big)
       -\frac{8}{l}\sqrt\mu
       -\frac{4}{l}\Big(e^{2Y/l}-\mu e^{-2Y/l}\Big)\bigg]\\
 &=\frac{1}{16\pi G}\int\rd^2x\,
 \bigg[\frac{8\mu}{l}e^{-2Y/l}-\frac{8}{l}\sqrt\mu\bigg]
 \xrightarrow[Y\to\infty]{}
 -\frac{8\sqrt\mu}{16\pi Gl}\int\rd^2x\,.
 \end{aligned}
 \label{SMgamma2cutoff}
\ee
The three terms in the first line are, respectively, the
on-shell bulk contribution at the cutoff, the interior
endpoint contribution, and the volume counterterm.
With no additional boundary action term at the interior endpoint, the renormalized
Lorentzian on-shell action takes the same form on both branches,
\be
 S_{\rm ren}
 =-\frac{8\sqrt{\mu}}{16\pi Gl}\int\rd^{2}x\,.
 \label{SMgamma2value}
\ee

\subsection{Riemannian Saddle Action}
\label{SMsub:action:R}

For the Riemannian branch, ${\mu=L_{+}L_{-}}$ and, at constant
${L_{\pm}>0}$, the interior locus ${e^{-2d}=0}$ is the Killing horizon.
With no additional boundary action term at the interior endpoint,
Eq.~\eqref{SMgamma2value} gives its
Lorentzian on-shell value.

\subsection{Non-Riemannian Saddle Action}
\label{SMsub:action:NR}

For the non-Riemannian branch, ${\mu=\half L_{+}L_{-}}$.  For constant
${L_{\pm}>0}$, its interior
endpoint occurs at ${q=1}$, where ${e^{-2d}=0}$ and
${e^{2y/l}=\sqrt{L_{+}L_{-}/2}}$; we neither assume that it is a smooth
cap nor exclude a smooth completion.  With no additional boundary action
term at the interior endpoint,
Eq.~\eqref{SMgamma2value} gives its Lorentzian on-shell value.
The on-shell Lagrangian density, boundary term~\eqref{SMgamma2flux},
and counterterm are all independent of $\cW$: \emph{the hair does not contribute to
the on-shell action} for arbitrary $W_{0}$ and $W_{1}$.  Together with its
absence from the covariant surface charges, this makes $W_{1}$
charge-neutral soft hair.  The Euclidean continuation, the assignment of
temperature and entropy, the choice of a common ensemble, and the relative weights
of the saddles in~\eqref{RGKPW} remain open questions.  Resolving them requires
specifying the interior boundary conditions and any
boundary action terms needed for a well-defined
variational principle.

\section{Linearized Dynamics and Holographic Renormalization}
\label{SMsec:linear}
\begingroup
\renewcommand{\arraystretch}{1.15}

We obtain one-point functions and connected two-point functions from the variation of the renormalized on-shell action. The chiral functions $L_\pm(x^\pm)$ and the NR hair $W_1(x^+,x^-)$ are arbitrary. We work at ${W_0=0}$; a fixed nonzero $W_0$ is restored by the finite Kalb--Ramond gauge transformation
\be
 \Omega_b=\begin{pmatrix}\mathbf 1&0\\ b&\mathbf 1\end{pmatrix}\,,
 \qquad \cH\longrightarrow\Omega_b\cH\Omega_b^T\,,
 \qquad b_{+-}=-\half W_0\,,\quad b_{y\pm}=0\,.
 \label{SMBtransform}
\ee
Here $b$ is closed and locally exact, ${b_{\mu\nu}=2\partial_{[\mu}\lambda_{\nu]}}$.
The transformation shifts ${\cW\to\cW-2b_{+-}}$, leaves $d$ unchanged,
and transports both the background and its vielbeins.  It changes the
boundary source and is therefore excluded from the fixed-source gauge
redundancies.

\subsection{Constrained Variations and Linearized EDFE}
\label{SMsub:variations}

The projectors ${P=(\cJ+\cH)/2}$ and ${\bar P=(\cJ-\cH)/2}$
obey ${P\delta\cH P=\bar P\delta\cH\bar P=0}$.  Hence the independent
variation and its reconstruction are
\be
h_{p\brq}:=\delta\cH_{MN}V^M{}_p\brV^N{}_{\brq}\,,\qquad
\delta\cH_{MN}=2V_{(M}{}^p\brV_{N)}{}^{\brq}h_{p\brq}\,.
\label{SMcosetreconstruction}
\ee
Up to local Lorentz rotations, the variations of the DFT vielbeins of
the saddle are
\be
\ba{ll}
\delta V_{Mp}=\half\brV_{M}{}^{\brq}h_{p\brq}\,,\qquad&\qquad
\delta \brV_{M\brq}=-\half V_{M}{}^{p}h_{p\brq}\,.
\ea
\label{SMframevariation}
\ee
The corresponding response convention is
${\delta\cL=2e^{-2d}(-h^{p\brq}K_{p\brq}+\delta d\,\To)}$,
up to field equations and total derivatives, as in~\eqref{RKdef}.
The flat metric in a light-cone basis has ${\eta_{\oplus\ominus}=-1}$ and
${\eta_{\mathrm y\mathrm y}=1}$, with ${\bar\eta=-\eta}$ and all
other independent entries zero.  The limiting DFT vielbeins are
\be
V^\infty_{Mp}=\begin{pmatrix}
\tfrac1{\sqrt2}&0&0\\0&0&0\\0&0&\tfrac1{\sqrt2}\\
0&-\sqrt2&0\\0&0&0\\0&0&\tfrac1{\sqrt2}
\end{pmatrix}\,,\qquad
\brV^\infty_{M\brp}=\begin{pmatrix}
0&0&0\\0&\tfrac1{\sqrt2}&0\\0&0&\tfrac1{\sqrt2}\\
0&0&0\\\sqrt2&0&0\\0&0&-\tfrac1{\sqrt2}
\end{pmatrix}\,.
\label{SMinfinityvielbein}
\ee
The columns carry lower flat indices in the order
$(\oplus,\ominus,\mathrm y)$ and
$(\bar\oplus,\bar\ominus,\bar{\mathrm y})$.
Radial gauge ${h_{p\bar{\mathrm y}}=h_{\mathrm y\brq}=0}$ leaves
five independent sources, ordered as
$(h^{(0)}_{\ominus\bar\ominus},h^{(0)}_{\oplus\bar\oplus},
h^{(0)}_{\ominus\bar\oplus},h^{(0)}_{\oplus\bar\ominus},\delta d^{(0)})$;
each couples with coefficient $+2$ to its operator in~\eqref{Mainfivebyfive}.

The DFT Christoffel connection is
\be
\begin{aligned}
\Gamma_{KMN}&=2(P\partial_KP\bar P)_{[MN]}
 +2(\bar P_{[M}{}^R\bar P_{N]}{}^S-P_{[M}{}^RP_{N]}{}^S)\partial_RP_{SK}\\
&\quad-\frac4{D-1}(\bar P_{K[M}\bar P_{N]}{}^R+P_{K[M}P_{N]}{}^R)
 [\partial_Rd+(P\partial^SP\bar P)_{[SR]}]\,,\qquad D=3\,.
\end{aligned}
\label{SMbackgroundconnection}
\ee
The curvature variations are expressed through the universal
$\Orth(D,D)$-symmetric, fully covariant box operator
$\Box$~\cite{Lee:2025fme}:
\be
\delta\So
=2\Box\delta d-\nabla_M\Omega^M
  +2(PS\bar P)^{MN}V_M{}^p\brV_N{}^{\brq}h_{p\brq}\,,
\label{SMfullscalarvariation}
\ee
\be
\delta(PS\bar P)_{MN}
=-\tfrac14\Box\!\left(V_M{}^p\brV_N{}^{\brq}h_{p\brq}\right)
  +P_M{}^{[R}\bar P_N{}^{S]}\nabla_R\Omega_S
 -\tfrac12(PS\bar P)_M{}^R\brV_R{}^{\brq}V_N{}^p h_{p\brq}
  +\tfrac12\brV_M{}^{\brq}V^{Rp}h_{p\brq}(PS\bar P)_{RN}\,,
\label{SMfullcurvaturevariation}
\ee
with
\be
\Omega_M
=e^{2d}\nabla_N\delta\!\left(e^{-2d}\cH^N{}_M\right)
=\left(V^{Np}\brV_M{}^{\brq}+V_M{}^p\brV^{N\brq}\right)\cD_Nh_{p\brq}-2\cH_M{}^N\partial_N\delta d\,.
\label{SMfullOmega}
\ee
The section condition gives ${\Delta+\bar\Delta=0}$, so
\be
\Box:=\Delta-\bar\Delta=2\Delta=-2\bar\Delta\,.
\label{SMboxdefinition}
\ee
For tensors, the derivative, curvature and connection terms together
ensure full covariance. In Riemannian variables, $\Box$ combines
Riemann curvature, $H$-flux and dilaton gradients.
Its mixed-tensor and scalar actions are
\be
\begin{aligned}
\Delta h_{p\brq}={}&\cD_r\cD^r h_{p\brq}
+2\mathfrak R_{\brq\bar s p r}h^{r\bar s} +2\left(\mathfrak R_{[pr]}
-\frac12\Gamma^{MN}{}_p\Gamma_{MNr}
-\Gamma^M{}_{pr}\cD_M\right)h^r{}_{\brq}\,,\\
\Box(P\delta\cH\bar P)_{MN}
={}&2V_M{}^p\brV_N{}^{\brq}\Delta h_{p\brq}\,.
\end{aligned}
\label{SMmixedboxdefinition}
\ee
\be
\Box\delta d=\cH^{MN}\nabla_M\nabla_N\delta d
=e^{2d}\partial_M\!\left(e^{-2d}\cH^{MN}\partial_N\delta d\right)\,.
\ee
Here $\cD$ includes both spin connections, and
\be
\mathfrak R^R{}_{SMN}
=\partial_M\Gamma_N{}^R{}_S-\partial_N\Gamma_M{}^R{}_S
+\Gamma_M{}^R{}_T\Gamma_N{}^T{}_S
-\Gamma_N{}^R{}_T\Gamma_M{}^T{}_S\,,\qquad
\mathfrak R_{MN}=\mathfrak R^R{}_{MRN}\,.
\label{SMboxcurvature}
\ee
Flat indices denote DFT vielbein projections.  We retain $\Omega_M$:
radial gauge does not impose the harmonic condition ${\Omega_M=0}$.

\subsubsection*{Linearized DFT Einstein Equations}

The DFT Einstein tensor is~\cite{Park:2025core}
\be
G_{MN}=4(PS\bar P)_{[MN]}-\frac12\cJ_{MN}\So\,.
\label{SMEinsteintensor}
\ee
For ${\Lambda_{\rm DFT}=-2/l^2}$, the equation
${G_{MN}=-\Lambda_{\rm DFT}\cJ_{MN}}$ is equivalent to
${(PS\bar P)_{MN}=0}$ and ${\So=2\Lambda_{\rm DFT}=-4/l^2}$.
At fixed cosmological constant, the curvature variations of
Ref.~\cite{Lee:2025fme} therefore give the linearized EDFE
\be
\begin{aligned}
0=\delta\So&=2\Box\delta d-\nabla_M\Omega^M\,,\qquad \Omega_M=\nabla^N\delta\cH_{NM}
 -2\cH_M{}^N\partial_N\delta d\,,\\
0=\delta(PS\bar P)_{MN}
&=-\frac14\Box(P\delta\cH\bar P)_{MN}
 +P_M{}^{[R}\bar P_N{}^{S]}\nabla_R\Omega_S\,.
\end{aligned}
\label{SMexactboxEDFE}
\ee
Here the tensor action of $\Box$ retains the curvature and connection
terms in~\eqref{SMmixedboxdefinition}. Only the explicit $PS\bar P$
terms in~\eqref{SMfullscalarvariation} and~\eqref{SMfullcurvaturevariation}
vanish on the saddle; $S_{MN}$ itself need not vanish.
\subsection{Boundary Variations and Counterterms}
\label{SMsub:momenta}

The bulk action is $S_{\Gamma^2}=(16\pi G)^{-1}\int\rd^3x
[\cL_{\Gamma^2}-2\Lambda_{\rm DFT}e^{-2d}]$, where
${\Lambda_{\rm DFT}=-2/l^2}$ and
\be
\cL_{\Gamma^{2}}=e^{-2d}\So-\partial_{K}\big(e^{-2d}B^{K}\big)=e^{-2d}
\big(P^{KM}P^{LN}-\brP^{KM}\brP^{LN}\big)
\big(\Gamma_{KM}{}^{J}\Gamma_{LNJ}-\Gamma_{KL}{}^{J}\Gamma_{NMJ}+\half\Gamma^{J}{}_{KL}\Gamma_{JMN}\big)\,,
\label{GammaDFT}
\ee
\be
\delta\cL_{\Gamma^2}=2e^{-2d}\left(h^{p\brq}S_{p\brq}-\delta d\,\So\right)
+\partial_{K}\left(h^{p\brq}e^{-2d}A^{K}_{p\brq}+
2\delta d\,e^{-2d}B^{K}\right)\,.
\label{variation}
\ee
The radial momenta are constructed from
\begin{subequations}
\label{defB}
\be
B^{K}:=2(P^{KM}P^{LN}-\brP^{KM}\brP^{LN})\Gamma_{LMN}
=4\cH^{KL}\partial_Ld-\partial_L\cH^{KL}\,,
\label{SMBdefinition}
\ee
\be
A^{K}_{p\brq}:=V^K{}_p\Gamma^L{}_{L\brq}
+\brV^K{}_{\brq}\Gamma^L{}_{Lp}
-\Gamma_p{}^K{}_{\brq}-\Gamma_{\brq}{}^K{}_p\,.
\label{SMAdefinition}
\ee
\end{subequations}
Including the cosmological term replaces $\So$ in the bulk variation
by $\So+4/l^2$.  On a saddle, the bulk terms vanish.  The volume counterterm is
\be
 S_{\rm ct}=-\frac{1}{16\pi G}\frac4l
 \int_{y=Y}\rd^2x\,e^{-2d}\,.
 \label{SMvolumecounterterm}
\ee
The resulting boundary variation at the ultraviolet cutoff is
\be
 \left.\delta(S_{\Gamma^2}+S_{\rm ct})\right|_{{\rm EOM},Y}
 =\frac{1}{16\pi G}\int_{y=Y}\rd^2x\,e^{-2d}
 \left[h^{p\brq}A^{y}_{p\brq}
 +2\delta d\left(B^y+\frac4l\right)\right]\,.
 \label{SMrenvariation}
\ee
In unit radial gauge, ${\cH^{yM}=\delta^M_y}$,
${B^y=4\partial_yd}$ and
${h^{p\brq}A^y_{p\brq}=\tfrac14\delta\cH^{AB}\partial_y\cH_{AB}}$.
Thus
\be
\left.\delta(S_{\Gamma^2}+S_{\rm ct})\right|_{{\rm EOM},Y}
=\frac1{16\pi G}\int_{y=Y}\rd^2x\,e^{-2d}
\left[\frac14\delta\cH^{AB}\partial_y\cH_{AB}
+8\delta d\left(\partial_yd+\frac1l\right)\right]\,.
\label{SMradialmetricvariation}
\ee
Indices are raised with the fixed $\cJ$.  At the cutoff, we use affine
dilaton and local coset coordinates, with ${\delta^{2}d=0}$ and
${P\delta^{2}\cH\bar P=0}$; the constraint fixes a same-chirality
quadratic completion orthogonal to $\partial_y\cH$.
The ultraviolet contribution to the second variation is
\be
\begin{aligned}
\left.\delta^{2}S_{\rm ren}\right|_{\rm UV}
=\lim_{Y\to\infty}\Bigg\{\frac{1}{16\pi G}\int_{y=Y}\rd^2x\,e^{-2d}
\Big[&\frac14\delta\cH^{AB}\partial_y\delta\cH_{AB}
+8\delta d\,\partial_y\delta d
-\frac12\delta d\,\delta\cH^{AB}\partial_y\cH_{AB}\\
&-16(\delta d)^2\left(\partial_yd+\frac1l\right)\Big]
+\delta^2S_{\rm ct}^{(2)}\Bigg\}\,.
\end{aligned}
\label{SMsecondvariation}
\ee
The last two terms inside the square bracket come from varying $e^{-2d}$.
The quadratic source counterterm $S_{\rm ct}^{(2)}$, given below,
supplements the volume counterterm already included.
The variations are extended to linearized bulk solutions.
Equation~\eqref{SMsecondvariation} gives the full second variation
when the interior boundary variation vanishes; otherwise that
contribution must be included.

Let $j_I$ denote the five ordered sources above, with conjugate
operators $\mathcal O_I$ in~\eqref{Mainfivebyfive}.
Once the state and interior conditions specify the on-shell action,
the connected two-point functions are
\be
\langle\mathcal O_I(x)\mathcal O_J(x')\rangle_c
=\left.\frac1{4i}\frac{\delta^2S_{\rm ren}}
{\delta j_I(x)\delta j_J(x')}\right|_{j=0}\,.
\label{SMRvacuumhessian}
\ee
The factor $1/4$ follows from the source couplings $+2j_I\mathcal O_I$.
All sources are retained until after differentiation, and the source
dependence of the bulk solutions is included. Below we read the
corresponding kernels from the second variation
in~\eqref{SMsecondvariation}.

\paragraph{Quadratic source counterterms.}
\label{SMsub:derivativect}
For general chiral $L_\pm$ and NR hair $W_1$, the divergences
involving boundary derivatives are canceled by
\be
\begin{aligned}
S_{{\rm ct},{\rm NR}}^{(2)}
=-\frac1{32\pi G}\int_{y=Y}\rd^2x\,\Bigg\{&
e^{2Y/l}\Big[
\frac l2 h_{\ominus\bar\ominus}\partial_+^2h_{\ominus\bar\oplus}
+\frac l2 h_{\oplus\bar\oplus}\partial_-^2h_{\ominus\bar\oplus}
-2l\delta d\,\partial_+\partial_-h_{\ominus\bar\oplus}
+\frac{l^3}{16}h_{\ominus\bar\oplus}\partial_+^2\partial_-^2h_{\ominus\bar\oplus}\Big]\\
&+2Yh_{\ominus\bar\ominus}
\Big(L_+\partial_++\frac12\partial_+L_+\Big)\partial_-h_{\ominus\bar\oplus}+2Yh_{\oplus\bar\oplus}
\Big(L_-\partial_-+\frac12\partial_-L_-\Big)\partial_+h_{\ominus\bar\oplus}\\
&+\frac{lY^2}{2}h_{\ominus\bar\oplus}
\Big[\Big(L_+\partial_++\frac12\partial_+L_+\Big)\partial_-^3
+\Big(L_-\partial_-+\frac12\partial_-L_-\Big)\partial_+^3\Big]
h_{\ominus\bar\oplus}\Bigg\}\,.
\end{aligned}
\label{SMderivativectNR}
\ee
On the Riemannian branch, the counterterm is
\be
\begin{aligned}
S_{{\rm ct},{\rm R}}^{(2)}
=S_{{\rm ct},{\rm NR}}^{(2)}
-\frac1{32\pi G}\int_{y=Y}\rd^2x\,
\Big[l^2Y\delta d\,\partial_+^2\partial_-^2h_{\ominus\bar\oplus}
+\frac{l^3Y^2}{16}h_{\ominus\bar\oplus}
\partial_+^3\partial_-^3h_{\ominus\bar\oplus}\Big]\,.
\end{aligned}
\label{SMderivativectR}
\ee
All fields here are evaluated at the cutoff.  In the Riemannian expression,
the NR counterterm polynomial is evaluated on Riemannian fluctuation fields,
and the volume counterterm~\eqref{SMvolumecounterterm} is retained.
These subtractions are quadratic about a fixed saddle; a universal
nonlinear completion that is intrinsic and DFT-covariant remains to be established.
The full cutoff fields, including their $e^{-2Y/l}$ terms,
must be substituted before taking the limit: the factors $e^{2Y/l}$
in the counterterms also produce finite contact terms.
These terms are retained in~\eqref{SMsecondvariation}.
Finite local counterterms change contact terms but not the
separated-point kernels considered below.

\paragraph{An intrinsic curvature candidate and its limitations.}
For comparison, let $\cH^\partial_{AB}$ and $d_\partial$ be the
induced tangential fields at ${y=Y}$ in radial gauge, with corresponding
projectors $P_\partial$ and $\bar P_\partial$.
Construct $S_{(0)}^\partial$ and
$(P_\partial S^\partial\bar P_\partial)_{AB}$ from the intrinsic
${D=2}$ DFT Christoffel connection using only tangential derivatives.
The $\mathbf{O}(D,D)$-symmetric box operator~\cite{Lee:2025fme} acts on a
weight-zero scalar $f$ as
${\Box_\partial f=e^{2d_\partial}\partial_A
(e^{-2d_\partial}\cH_\partial^{AB}\partial_B f)}$.
With indices raised by $\cJ_\partial$, consider
\be
\resizebox{\linewidth}{!}{$\displaystyle
S_{\rm ct}^{\partial} =-\frac1{32\pi G}\int_{y=Y}\rd^2x\,e^{-2d_\partial} \Bigg[\frac8l+lS_{(0)}^\partial +\frac{l^3}{16}(S_{(0)}^\partial)^2 -\frac{l^2Y}{8}\Big((S_{(0)}^\partial)^2 +8(P_\partial S^\partial\bar P_\partial)_{AB} (P_\partial S^\partial\bar P_\partial)^{AB}\Big) +\frac{l^3Y^2}{32}S_{(0)}^\partial\Box_\partial S_{(0)}^\partial \Bigg]\,.
$}
\label{SMboundarycurvaturecandidate}
\ee
These are intrinsic boundary curvatures, not restrictions of the bulk
curvatures. The volume counterterm~\eqref{SMvolumecounterterm} is
included.
At quadratic order about the non-Riemannian vacuum,
\eqref{SMboundarycurvaturecandidate} reproduces all divergent and finite terms
of the volume counterterm~\eqref{SMvolumecounterterm} plus the source
counterterm~\eqref{SMderivativectNR} as ${Y\to\infty}$, modulo tangential
total derivatives. On the Riemannian vacuum, a finite local adjustment is
needed to match the contact-term convention.

For generic constant nonzero $L_\pm$,
\eqref{SMboundarycurvaturecandidate} leaves uncanceled divergences on both
branches. The matching above is therefore restricted to quadratic order
about the vacua; a general nonlinear counterterm remains to be established.
For general states we retain
\eqref{SMderivativectNR} and \eqref{SMderivativectR}.

\subsection{Riemannian Branch: Linearized Solutions}
\label{SMsub:linear:R}
\label{SMsub:interior}

We substitute the constrained fluctuation
${\delta\cH_{MN}=2V_{(M}{}^p\brV_{N)}{}^{\brq}h_{p\brq}}$
into the linearized EDFE, using the exact saddle vielbeins.  For the Riemannian branch, these are
\be
\begin{gathered}
e=\begin{pmatrix}1&-L_+&0\\-e^{-2y/l}L_-&e^{2y/l}&0\\0&0&1\end{pmatrix}\,,\qquad
\bar e=\begin{pmatrix}e^{2y/l}&-e^{-2y/l}L_+&0\\-L_-&1&0\\0&0&1\end{pmatrix}\,,\\
V=\frac1{\sqrt2}\binom{e^{-T}}{e\eta+Be^{-T}}\,,\qquad
\brV=\frac1{\sqrt2}\binom{\bar e^{-T}}{\bar e\bar\eta+B\bar e^{-T}}\,,\qquad
d=-\frac yl-\frac12\ln(1-L_+L_-e^{-4y/l})\,.
\end{gathered}
\label{SMexactRdata}
\ee
Here $B$ is given in~\eqref{Rfields}, and the DFT 
vielbein matrices carry lower flat indices.

\subsubsection*{Vacuum Response from Interior Regularity}

An explicit interior prescription is available on the Riemannian
vacuum ${L_\pm=W_0=0}$.  We choose the
Euclidean-regular Poincar\'e vacuum on the covering space, fixing its chiral
homogeneous modes and constant mode.  For nonzero modes, define
\be
\delta d^{(2)}
=\frac{l^2}{8}\left(\partial_-^2h^{(0)}_{\oplus\bar\oplus}
+\partial_+^2h^{(0)}_{\ominus\bar\ominus}
-4\partial_+\partial_-\delta d^{(0)}\right)
-\frac{l^4}{32}\partial_+^2\partial_-^2h^{(0)}_{\ominus\bar\oplus}\,.
\label{SMRvacuumdilatoncoefficient}
\ee
We display the near-boundary expansion of the full radial solution as
\be
\begin{aligned}
h_{\oplus\bar\oplus}
&=h^{(0)}_{\oplus\bar\oplus}
+e^{-2y/l}\left(2\partial_-^{-1}\partial_+\delta d^{(2)}
-\frac{l^4}{16}\partial_+^3\partial_-h^{(0)}_{\ominus\bar\oplus}\right)
-\frac l2y\left(\partial_+^2+\frac{l^2}{4}e^{-2y/l}\partial_+^3\partial_-\right)h^{(0)}_{\ominus\bar\oplus}\,,\\
h_{\ominus\bar\ominus}
&=h^{(0)}_{\ominus\bar\ominus}
+e^{-2y/l}\left(2\partial_+^{-1}\partial_-\delta d^{(2)}
-\frac{l^4}{16}\partial_+\partial_-^3h^{(0)}_{\ominus\bar\oplus}\right)
-\frac l2y\left(\partial_-^2+\frac{l^2}{4}e^{-2y/l}\partial_+\partial_-^3\right)h^{(0)}_{\ominus\bar\oplus}\,,\\
h_{\ominus\bar\oplus}
&=h^{(0)}_{\ominus\bar\oplus}+e^{-2y/l}\left(4\delta d^{(0)}-\frac{l^2}{4}\partial_+\partial_-h^{(0)}_{\ominus\bar\oplus}\right)
-\frac l2y e^{-2y/l}\partial_+\partial_-h^{(0)}_{\ominus\bar\oplus}\,,\\
h_{\oplus\bar\ominus}
&=h^{(0)}_{\oplus\bar\ominus}+e^{-2y/l} H_{\rm R}
+\frac{l^2}{8}y^2\partial_+^2\partial_-^2h^{(0)}_{\ominus\bar\oplus}
-\frac4l y\delta d^{(2)}
-\frac{l^3}{4}y\left(\partial_+^2\partial_-^2
+\frac{l^2}{4}e^{-2y/l}\partial_+^3\partial_-^3\right)h^{(0)}_{\ominus\bar\oplus}\,,\\
\delta d
&=\delta d^{(0)}-\frac l8y\partial_+\partial_-h^{(0)}_{\ominus\bar\oplus}
+e^{-2y/l}\left(\delta d^{(2)}-\frac{l^3}{16}y\partial_+^2\partial_-^2h^{(0)}_{\ominus\bar\oplus}\right)\,.
\end{aligned}
\label{SMRvacuumsolution}
\ee
Terms of order $\cO(e^{-4y/l}\operatorname{poly}(y))$ are omitted.
The response $H_{\rm R}$ is given by
\be
H_{\rm R}
=\frac{l^6}{16}\partial_+^3\partial_-^3
\left[C_\mu+\frac12\ln\!\left(\frac{\partial_+\partial_-}{\mu^2}\right)\right]h^{(0)}_{\ominus\bar\oplus}\,,
\qquad
C_\mu=\ln\!\left(\frac{l\mu}{\sqrt2}\right)+\gamma_E-\frac32\,.
\label{SMRvacuumresponse}
\ee
The logarithm is defined spectrally in Euclidean momentum space, with $\partial_+\partial_-$
having eigenvalue $k^2/2$; the arbitrary inverse-length scale $\mu$
cancels.  Here $\gamma_E$ is the Euler--Mascheroni constant.
Lorentzian correlators follow by analytic continuation.

\subsubsection*{General Chiral Backgrounds}

For arbitrary chiral $L_\pm$ and the five independent sources,
we use the near-boundary expansion
\be
h_{a\brb}=\sum_{n=0,2}\sum_{k=0}^{2}h^{(n,k)}_{a\brb}(x)\,y^ke^{-ny/l}
+\cO(e^{-4y/l}\operatorname{poly}(y))\,,\qquad
\delta d=\sum_{n=0,2}\sum_{k=0}^{2}\delta d^{(n,k)}(x)\,y^ke^{-ny/l}
+\cO(e^{-4y/l}\operatorname{poly}(y))\,.
\label{SMradialansatz}
\ee
Here ${h^{(0,0)}=h^{(0)}}$ and ${\delta d^{(0,0)}=\delta d^{(0)}}$.
The other nonzero coefficients are
\be
\begin{aligned}
h_{\oplus\bar\oplus}^{(0,1)}&=-\frac l2\partial_+^2h^{(0)}_{\ominus\bar\oplus}\,,\qquad h_{\oplus\bar\oplus}^{(2,1)}=-\frac l8
\left[l^2\partial_+^3-4L_+\partial_+-4(\partial_+L_+)\right]\partial_-h^{(0)}_{\ominus\bar\oplus}\,,\\
h_{\oplus\bar\oplus}^{(2,0)}&=b_{++}+\omega_+
+L_+\left(4\delta d^{(0)}-4f_0-\frac{l^2}{4}\partial_+\partial_-h^{(0)}_{\ominus\bar\oplus}\right)
 -\frac{l^2}{4}\left[\frac{3l^2}{4}\partial_+^3-6L_+\partial_+
-4(\partial_+L_+)\right]\partial_-h^{(0)}_{\ominus\bar\oplus}\,.
\end{aligned}
\label{SMRstresspluscoefficients}
\ee
\be
\begin{aligned}
h_{\ominus\bar\ominus}^{(0,1)}&=-\frac l2\partial_-^2h^{(0)}_{\ominus\bar\oplus}\,,\qquad h_{\ominus\bar\ominus}^{(2,1)}=-\frac l8
\left[l^2\partial_-^3-4L_-\partial_--4(\partial_-L_-)\right]\partial_+h^{(0)}_{\ominus\bar\oplus}\,,\\
h_{\ominus\bar\ominus}^{(2,0)}&=b_{--}+\omega_-
+L_-\left(4\delta d^{(0)}-4f_0-\frac{l^2}{4}\partial_+\partial_-h^{(0)}_{\ominus\bar\oplus}\right)
 -\frac{l^2}{4}\left[\frac{3l^2}{4}\partial_-^3-6L_-\partial_-
-4(\partial_-L_-)\right]\partial_+h^{(0)}_{\ominus\bar\oplus}\,.
\end{aligned}
\label{SMRstressminuscoefficients}
\ee
\be
\begin{aligned}
h_{\ominus\bar\oplus}^{(2,0)}&=4\delta d^{(0)}-4f_0
-\frac{l^2}{4}\partial_+\partial_-h^{(0)}_{\ominus\bar\oplus}\,,&
h_{\ominus\bar\oplus}^{(2,1)}&=-\frac l2\partial_+\partial_-h^{(0)}_{\ominus\bar\oplus}\,.
\end{aligned}
\label{SMRtypecoefficients}
\ee
\be
\begin{aligned}
h_{\oplus\bar\ominus}^{(0,1)}&=-\frac l2\left(
\partial_-^2h^{(0)}_{\oplus\bar\oplus}+\partial_+^2h^{(0)}_{\ominus\bar\ominus}
-4\partial_+\partial_-\delta d^{(0)}+\frac{l^2}{4}\partial_+^2\partial_-^2h^{(0)}_{\ominus\bar\oplus}\right)\,,
\qquad h_{\oplus\bar\ominus}^{(0,2)}=\frac{l^2}{8}\partial_+^2\partial_-^2h^{(0)}_{\ominus\bar\oplus}\,,
\qquad h_{\oplus\bar\ominus}^{(2,0)}=H_{\rm R}\,,\\
h_{\oplus\bar\ominus}^{(2,1)}&=-\frac{l^3}{4}
\left[\frac{l^2}{8}\partial_+^3-\frac32L_+\partial_+-(\partial_+L_+)\right]\partial_-^3h^{(0)}_{\ominus\bar\oplus}
-\frac{l^3}{4}\left[\frac{l^2}{8}\partial_-^3-\frac32L_-\partial_--(\partial_-L_-)\right]\partial_+^3h^{(0)}_{\ominus\bar\oplus}\\
&\quad-\frac l2\left[5L_+L_-\partial_+\partial_-
+3(\partial_-L_-)L_+\partial_++3(\partial_+L_+)L_-\partial_-
+2(\partial_+L_+)(\partial_-L_-)\right]h^{(0)}_{\ominus\bar\oplus}\,.
\end{aligned}
\label{SMRhaircoefficients}
\ee
\be
\begin{aligned}
\delta d^{(0,1)}&=-\frac l8\partial_+\partial_-h^{(0)}_{\ominus\bar\oplus}\,,\qquad \delta d^{(2,1)}=-\frac{l^3}{16}\partial_+^2\partial_-^2h^{(0)}_{\ominus\bar\oplus}\,,\\
\delta d^{(2,0)}&=\frac{l^2}{8}\left(
\partial_-^2h^{(0)}_{\oplus\bar\oplus}+\partial_+^2h^{(0)}_{\ominus\bar\ominus}
-4\partial_+\partial_-\delta d^{(0)}\right)
 -\frac{l^2}{32}\left(l^2\partial_+^2\partial_-^2-4L_+\partial_-^2-4L_-\partial_+^2\right)h^{(0)}_{\ominus\bar\oplus}\,.
\end{aligned}
\label{SMRdilatoncoefficients}
\ee
Here $f_0$ is constant.  The response $H_{\rm R}$ contains an explicitly
determined source-dependent part and a remaining scalar integration function,
as displayed below.
For the Euclidean-regular Poincar\'e vacuum specified above,
${f_0=0}$ and the additional chiral homogeneous freedom in
$\lambda_\pm^{\rm R}$ below is fixed by the vacuum and zero-mode prescription.
The auxiliary coefficients are
\be
\begin{aligned}
a_{+-}&=2\delta d^{(0)}-2f_0-\frac{l^2}{8}\partial_+\partial_-h^{(0)}_{\ominus\bar\oplus}\,,\qquad a_{++}=h^{(0)}_{\oplus\bar\oplus}-L_+h^{(0)}_{\ominus\bar\oplus}\,,\qquad
a_{--}=h^{(0)}_{\ominus\bar\ominus}-L_-h^{(0)}_{\ominus\bar\oplus}\,,\\
b_{+-}&=-L_-a_{++}-L_+a_{--}
+\frac{l^2}{4}\left(\partial_+^2a_{--}+\partial_-^2a_{++}-2\partial_+\partial_-a_{+-}\right)\,,\\
b_{++}&=\partial_-^{-1}\left[\partial_+b_{+-}+L_-\partial_+a_{++}-L_+\partial_+a_{--}\right]+\lambda_+^{\rm R}(x^+)\,,\\
b_{--}&=\partial_+^{-1}\left[\partial_-b_{+-}-L_-\partial_-a_{++}+L_+\partial_-a_{--}\right]+\lambda_-^{\rm R}(x^-)\,,\\
\omega_+&=\frac12\partial_-^{-1}\left\{L_-
\left[l^2\partial_+^3-8L_+\partial_+-4(\partial_+L_+)\right]h^{(0)}_{\ominus\bar\oplus}\right\}\,,\\
\omega_-&=\frac12\partial_+^{-1}\left\{L_+
\left[l^2\partial_-^3-8L_-\partial_--4(\partial_-L_-)\right]h^{(0)}_{\ominus\bar\oplus}\right\}\,.
\end{aligned}
\label{SMRintegratedstress}
\ee
In particular, the off-diagonal response coefficient is
\be
\begin{aligned}
H_{\rm R}={}&L_-(b_{++}+\omega_+)+L_+(b_{--}+\omega_-)
+L_+L_-\left(4\delta d^{(0)}-4f_0
-\frac{l^2}{4}\partial_+\partial_-h^{(0)}_{\ominus\bar\oplus}\right)
+F_{\rm R}\\
&-\frac{l^4}{48}\left[\frac{7l^2}{8}\partial_+^3
-2L_+\partial_+-7(\partial_+L_+)\right]\partial_-^3h^{(0)}_{\ominus\bar\oplus}
-\frac{l^4}{48}\left[\frac{7l^2}{8}\partial_-^3
-2L_-\partial_--7(\partial_-L_-)\right]\partial_+^3h^{(0)}_{\ominus\bar\oplus}\\
&+\frac{l^2}{12}\left[20L_+L_-\partial_+\partial_-
+(\partial_-L_-)L_+\partial_++(\partial_+L_+)L_-\partial_-
-7(\partial_+L_+)(\partial_-L_-)\right]h^{(0)}_{\ominus\bar\oplus}\,.
\end{aligned}
\label{SMRoffdiagonalresponse}
\ee
The remaining function $F_{\rm R}(x^+,x^-)$ is not fixed by the asymptotic
equations.  Its source dependence requires an interior prescription.  The explicit $b_{\pm\pm}+\omega_\pm$ terms already determine
particular nonlocal responses.
Coefficients not listed vanish through this order.

\subsubsection*{Nonlocal Kernels}

Define
\be
\mathbb K_\pm=
\left[\frac{l^2}{4}\partial_\pm^3-2L_\pm\partial_\pm
-(\partial_\pm L_\pm)\right]\partial_\mp^{-1}\,.
\label{SMRnonlocalkernel}
\ee
The inverse derivatives require a choice of homogeneous modes.
Moving $\partial_\mp^{-1}$ to the left changes only chiral
homogeneous terms, which are included in $\lambda_\pm^{\rm R}$.
Modulo local source terms, \eqref{SMRintegratedstress} then gives
\be
b_{\pm\pm}+\omega_\pm
=\mathbb K_\pm\big(h^{(0)}_{\mp\bar\mp}+L_\mp h^{(0)}_{\ominus\bar\oplus}\big)
+\lambda_\pm^{\rm R}\,.
\label{SMRnonlocalresponse}
\ee
Here $\pm\bar\pm$ denotes $\oplus\bar\oplus$ or $\ominus\bar\ominus$,
and $\mp\bar\mp$ denotes the other pair.
Substitution in~\eqref{SMsecondvariation}, with the constant mode held
fixed, gives the nonlocal ultraviolet contribution
\be
\left.\delta^2S_{\rm ren}\right|_{{\rm UV},{\rm nl}}
=\frac1{16\pi Gl}\int\rd^2x\,
\Bigg[\sum_\pm\big(h^{(0)}_{\mp\bar\mp}+L_\mp h^{(0)}_{\ominus\bar\oplus}\big)
\Big\{\mathbb K_\pm\big(h^{(0)}_{\mp\bar\mp}+L_\mp h^{(0)}_{\ominus\bar\oplus}\big)
+\lambda_\pm^{\rm R}\Big\}
 +h^{(0)}_{\ominus\bar\oplus}F_{\rm R}\Bigg]_{\rm nl}\,.
\label{SMRnonlocalaction}
\ee
The subscript ${\rm nl}$ retains the nonlocal source dependence.
In particular, $h^{(0)}_{\ominus\bar\oplus}F_{\rm R}$ must be kept: it contains the logarithmic
hair response~\eqref{SMRvacuumresponse} on the Riemannian vacuum.
The source dependence of $F_{\rm R}$ and $\lambda_\pm^{\rm R}$ is
fixed by the state and interior conditions.

The source bilinear in~\eqref{SMRnonlocalaction} determines the
two-point kernels through~\eqref{SMRvacuumhessian}, with the common
coefficient $1/(64\pi iGl)$. The integrations by parts are understood
with the chosen homogeneous-mode prescription; its additional
contributions are included in $\mathcal R_{\rm R}$.

To evaluate the position-space kernels, we use the joint Lorentzian
prescription with ${\Delta_\pm=x^\pm-x'^\pm}$:
\be
\delta^2(x-x')=\frac1{2\pi i}\partial_+\partial_-
\ln(-\Delta_+\Delta_-+i\epsilon)\,.
\label{SMLorentziandeltaconvention}
\ee
Differentiation precedes the separated-point limit, in which
\be
\partial_\mp^{-1}\delta^2\longrightarrow\frac1{2\pi i\Delta_\pm}\,,
\qquad
\ln(\partial_+\partial_-/\mu^2)\delta^2
\longrightarrow\frac{i}{2\pi\Delta_+\Delta_-}\,.
\label{SMnonlocalkernelprescription}
\ee
Using ${\partial_\pm\Delta_\pm^{-1}=-\Delta_\pm^{-2}}$ and
${\partial_\pm^3\Delta_\pm^{-1}=-6\Delta_\pm^{-4}}$, we obtain
$A_\pm^{\rm R}$ in~\eqref{Rcorrelators} and
${M_\pm^{\rm R}(x,x')=L_\mp(x^\mp)A_\pm^{\rm R}(x,x')}$.
These are the particular kernels; the source dependence of
$F_{\rm R}$ and $\lambda_\pm^{\rm R}$ must also be differentiated
for the complete correlators. The displayed kernels apply away from
contact and light-cone singularities, before imposing the spatial identification.
Additional source dependence of the integration functions contributes
to $\mathcal R_s$ in~\eqref{Mainfivebyfive}.

For the Euclidean-regular Poincar\'e vacuum specified above,
\eqref{SMRvacuumresponse} fixes the hair contribution in
\eqref{SMRnonlocalaction}. Applying~\eqref{SMRvacuumhessian} and
\eqref{SMnonlocalkernelprescription}, the nonzero separated-point
vacuum correlators are
\be
\langle\KCFT_{\pm\bar\pm}(x)\KCFT_{\pm\bar\pm}(x')\rangle_{{\rm R},c}
=\frac{3l}{256\pi^2G\,\Delta_\pm^4}
=\frac1{(8\pi)^2}\frac{c_{\rm charge}^{\rm R}/2}{\Delta_\pm^4}\,,
\qquad c_{\rm charge}^{\rm R}=\frac{3l}{2G}\,,
\label{SMRtwopt}
\ee
\be
\langle\KCFT_{\oplus\bar\ominus}(x)
\KCFT_{\oplus\bar\ominus}(x')\rangle_{{\rm R},c}
=\frac{9l^5}{1024\pi^2G}
\frac1{[-\Delta_+\Delta_-+i\epsilon]^4}\,.
\label{SMRvacuumhaircorrelator}
\ee
Finite local counterterms affect only the contact terms.

\subsection{Non-Riemannian Vielbeins and Linearized Solutions}
\label{SMsub:linear:NR}

The exact three-dimensional non-Riemannian DFT vielbeins are
given below in the coordinate order fixed above, with upper flat
indices and the projector conventions of Sec.~\ref{SMsub:variations}.
The functions $\chi$ and $\sigma$ are defined in~\eqref{NRvariables}.
\be
 V_{M}{}^{p}=
 \begin{pmatrix}
 0&-\tfrac{1}{\sqrt2}\cosh\tfrac{\chi}{2}&0\\
 0&-\tfrac{1}{\sqrt2}e^{\sigma}\sinh\tfrac{\chi}{2}&0\\
 0&0&\tfrac{1}{\sqrt2}\\
 \sqrt2\cosh\tfrac{\chi}{2}&\tfrac{\cW}{2\sqrt2}e^{\sigma}\sinh\tfrac{\chi}{2}&0\\
 -\sqrt2e^{-\sigma}\sinh\tfrac{\chi}{2}&-\tfrac{\cW}{2\sqrt2}\cosh\tfrac{\chi}{2}&0\\
 0&0&\tfrac{1}{\sqrt2}
 \end{pmatrix}\,,
 \quad
 \brV_{M}{}^{\brp}=
 \begin{pmatrix}
 \tfrac{1}{\sqrt2}e^{-\sigma}\sinh\tfrac{\chi}{2}&0&0\\
 \tfrac{1}{\sqrt2}\cosh\tfrac{\chi}{2}&0&0\\
 0&0&-\tfrac{1}{\sqrt2}\\
 -\tfrac{\cW}{2\sqrt2}\cosh\tfrac{\chi}{2}&-\sqrt2e^{\sigma}\sinh\tfrac{\chi}{2}&0\\
 \tfrac{\cW}{2\sqrt2}e^{-\sigma}\sinh\tfrac{\chi}{2}&\sqrt2\cosh\tfrac{\chi}{2}&0\\
 0&0&\tfrac{1}{\sqrt2}
 \end{pmatrix}\,.
 \label{SMvielbein}
\ee
The DFT vielbein relations are
\be
 V_{M}{}^{p}V_{Np}=P_{MN}\,,
 \qquad
 \brV_{M}{}^{\bar p}\brV_{N\bar p}=\bar P_{MN}\,,
 \qquad
 V^{M}{}_{p}\brV_{M\bar q}=0\,,
 \qquad
 P_{MN}+\brP_{MN}=\cJ_{MN}\,.
 \label{SMvielbeincheck}
\ee
Equation~\eqref{SMvielbein} therefore reconstructs ${\cH_{MN}=P_{MN}-\brP_{MN}}$.
We use the exact DFT vielbeins~\eqref{SMvielbein}, lowering their displayed
flat indices with $\eta$ and $\bar\eta$ before projecting the fluctuations.
With the same five sources, the solution through order $e^{-2y/l}$ is
\be
\begin{aligned}
h_{\oplus\bar\oplus}&=h^{(0)}_{\oplus\bar\oplus}+e^{-2y/l} h^{(2)}_{\oplus\bar\oplus}
-\frac l2y\left[\partial_+^2h^{(0)}_{\ominus\bar\oplus}-e^{-2y/l}\partial_+(L_+\partial_-h^{(0)}_{\ominus\bar\oplus})\right]\,,\\
h_{\ominus\bar\ominus}&=h^{(0)}_{\ominus\bar\ominus}+e^{-2y/l} h^{(2)}_{\ominus\bar\ominus}
-\frac l2y\left[\partial_-^2h^{(0)}_{\ominus\bar\oplus}-e^{-2y/l}\partial_-(L_-\partial_+h^{(0)}_{\ominus\bar\oplus})\right]\,,
\qquad h_{\ominus\bar\oplus}=h^{(0)}_{\ominus\bar\oplus}+e^{-2y/l} c_{\rm NR}\,,\\
h_{\oplus\bar\ominus}&=h^{(0)}_{\oplus\bar\ominus}+e^{-2y/l} H_{\rm NR}
+\frac{l^2}{8}y^2\partial_+^2\partial_-^2h^{(0)}_{\ominus\bar\oplus}
-\frac l2y\left[\partial_-^2h^{(0)}_{\oplus\bar\oplus}
+\partial_+^2h^{(0)}_{\ominus\bar\ominus}-4\partial_+\partial_-\delta d^{(0)}
+\frac{l^2}{4}\partial_+^2\partial_-^2h^{(0)}_{\ominus\bar\oplus}-\frac2l e^{-2y/l}\mathcal U\right]\,,\\
\delta d&=\delta d^{(0)}+\frac{l^2}{8}e^{-2y/l}(L_-\partial_+^2+L_+\partial_-^2)h^{(0)}_{\ominus\bar\oplus}
-\frac l8y\partial_+\partial_-h^{(0)}_{\ominus\bar\oplus}\,.
\end{aligned}
\label{SMNRcompactsolution}
\ee
The remainder in every line is $\cO(e^{-4y/l}\operatorname{poly}(y))$.
$H_{\rm NR}(x^+,x^-)$ is free at this order and $c_{\rm NR}$ is constant, while
\be
\begin{aligned}
h^{(2)}_{\oplus\bar\oplus}
&=\frac{l^2}{4}\partial_-^{-1}\left[L_-\partial_+^3+4L_+\partial_+\partial_-^2
+3(\partial_+L_+)\partial_-^2\right]h^{(0)}_{\ominus\bar\oplus}+4L_+\delta d^{(0)}\\
&\quad-2L_+\partial_-^{-1}\partial_+h^{(0)}_{\ominus\bar\ominus}
-(\partial_+L_+)\partial_-^{-1}h^{(0)}_{\ominus\bar\ominus}
-\frac14\partial_-^{-1}\left[2W_1\partial_++(\partial_+W_1)\right]h^{(0)}_{\ominus\bar\oplus}
+\lambda_+^{\rm NR}(x^+)\,,\\
h^{(2)}_{\ominus\bar\ominus}
&=\frac{l^2}{4}\partial_+^{-1}\left[L_+\partial_-^3+4L_-\partial_+^2\partial_-
+3(\partial_-L_-)\partial_+^2\right]h^{(0)}_{\ominus\bar\oplus}+4L_-\delta d^{(0)}\\
&\quad-2L_-\partial_+^{-1}\partial_-h^{(0)}_{\oplus\bar\oplus}
-(\partial_-L_-)\partial_+^{-1}h^{(0)}_{\oplus\bar\oplus}
-\frac14\partial_+^{-1}\left[2W_1\partial_-+(\partial_-W_1)\right]h^{(0)}_{\ominus\bar\oplus}
+\lambda_-^{\rm NR}(x^-)\,.
\end{aligned}
\label{SMNRintegratedstress}
\ee
\be
\begin{aligned}
\mathcal U={}&\frac l2\left[2L_+\partial_+\partial_-h^{(0)}_{\ominus\bar\ominus}
+(\partial_+L_+)\partial_-h^{(0)}_{\ominus\bar\ominus}
+2L_-\partial_+\partial_-h^{(0)}_{\oplus\bar\oplus}
+(\partial_-L_-)\partial_+h^{(0)}_{\oplus\bar\oplus}\right]\\
&-2l(L_+\partial_-^2+L_-\partial_+^2)\delta d^{(0)}
+\frac l2\left(\partial_-^2h^{(2)}_{\oplus\bar\oplus}
+\partial_+^2h^{(2)}_{\ominus\bar\ominus}\right)\\
&-\frac{l^3}{4}\partial_+(L_+\partial_-^3h^{(0)}_{\ominus\bar\oplus})
-\frac{l^3}{4}\partial_-(L_-\partial_+^3h^{(0)}_{\ominus\bar\oplus})
-\frac l8W_1\partial_+\partial_-h^{(0)}_{\ominus\bar\oplus}\,.
\end{aligned}
\label{SMNRcompactU}
\ee
With $c_{\rm NR}$ held fixed and the source dependence of
$H_{\rm NR}$ retained, substitution in~\eqref{SMsecondvariation}
gives the nonlocal contribution to the second variation,
with the normalization in~\eqref{SMRvacuumhessian}:
\be
\left.\delta^2S_{\rm ren}\right|_{{\rm UV},{\rm nl}}
=\frac1{16\pi Gl}\int\rd^2x\,
\left[h^{(0)}_{\ominus\bar\ominus}h^{(2)}_{\oplus\bar\oplus}
+h^{(0)}_{\oplus\bar\oplus}h^{(2)}_{\ominus\bar\ominus}
+h^{(0)}_{\ominus\bar\oplus}H_{\rm NR}\right]_{\rm nl}\,.
\label{SMNRnonlocalaction}
\ee
Applying~\eqref{SMRvacuumhessian} with the prescription
\eqref{SMnonlocalkernelprescription} gives the particular NR kernels
in~\eqref{Rcorrelators}. The remaining source dependence of
$H_{\rm NR}$ and $\lambda_\pm^{\rm NR}$ contributes to
$\mathcal R_{\rm NR}$ in~\eqref{Mainfivebyfive}.
\endgroup

\section{Worldsheet Actions and Vertex Operators}
\label{SMsec:worldsheet}

We use a Lorentzian worldsheet throughout, retaining the target-space
time coordinate on both branches.  With real null coordinates
${\sigma^\pm=(\sigma^0\pm\sigma^1)/\sqrt2}$, our orientation is
${\epsilon^{01}=1}$ and ${\epsilon^{+-}=-1}$.
In conformal gauge, ${\sqrt{-h}\,h^{+-}=\sqrt{-h}\,h^{-+}=-1}$ and
${h^{++}=h^{--}=0}$.  We write
${\partial=\partial/\partial\sigma^+}$ and
${\bar\partial=\partial/\partial\sigma^-}$; bars distinguish the two
real chiral sectors.  The generalized metric couples to the string
through the doubled-yet-gauged action~\cite{Lee:2013hma,Ko:2015rha,Morand:2017fnv},
\be
 S_{\rm dyg}
 =\frac{1}{4\pi\alpha'}\!\int\!\rd^{2}\sigma
 \Big[-\half\sqrt{-h}\,h^{\alpha\beta}
 D_{\alpha}x^{M}D_{\beta}x^{N}\,\cH_{MN}
 -\epsilon^{\alpha\beta}D_{\alpha}x^{M}\mathfrak{a}_{\beta M}\Big]\,,
 \label{SMdyg}
\ee
where ${D_\alpha x^M=\partial_\alpha x^M-\mathfrak a_\alpha{}^M}$ and
the auxiliary potential gauges the dual directions, with
${\mathfrak a_\alpha{}^M\partial_M=0}$.
On the physical section ${\tilde\partial^\mu=0}$ it has the form
${\mathfrak a_\alpha{}^M=(\mathfrak a_{\alpha\mu};0)}$.
Absorbing the dual-coordinate derivative gives
\be
 \widehat{\mathfrak{a}}_{\alpha\mu}
 :=\mathfrak{a}_{\alpha\mu}-\partial_{\alpha}\tilde x_{\mu}\,,
 \qquad
 D_{\alpha}x^{M}
 =\big(-\widehat{\mathfrak{a}}_{\alpha\mu};
       \partial_{\alpha}x^{\mu}\big)\,.
 \label{SMphysicalsectionA}
\ee
We drop the hat below.  This replacement changes the topological term
only by a worldsheet total derivative.  All auxiliary components can
be eliminated quadratically on the Riemannian branch.  On the
non-Riemannian branch, two auxiliary combinations in the $x^\pm$ sector
instead impose chirality constraints.  The generalized Fradkin--Tseytlin coupling
has the undoubled coefficient~\cite{Blair:2020gnh}; its Lorentzian
sign and stress-tensor contribution are specified below.

\subsection{Riemannian Sigma Model and Long Strings}
\label{SMsub:worldsheet:R}

Eliminating the auxiliary potential in~\eqref{SMdyg} gives the
NS--NS sigma model for~\eqref{Rfields}.  In the above orientation its
conformal-gauge Lagrangian is
${(2\pi\alpha')^{-1}(g_{\mu\nu}-B_{\mu\nu})
\partial x^\mu\bar\partial x^\nu}$.
Writing ${F=e^{2y/l}+L_+L_-e^{-2y/l}}$, the $x^\pm$ block of
${E=g-B}$ has ${E_{+-}=0}$, ${E_{-+}=-2F}$ and
${E_{\pm\pm}=2L_\pm}$.  The flux therefore cancels the $\partial x^+\bar\partial x^-$
coupling exactly.  For ${F\neq0}$, the remaining mixed coupling admits the
first-order representation
\be
L_{\rm R}^{(1)}=\frac{1}{2\pi\alpha'}\bigg[
\partial y\,\bar\partial y
+2L_+\,\partial x^+\bar\partial x^+
+2L_-\,\partial x^-\bar\partial x^-
+\beta\,\bar\partial x^+
+\brbeta\,\partial x^-
+\frac{\beta\brbeta}{2F}\bigg]\,.
\label{SMRfirstorder}
\ee
The equations ${\beta=-2F\,\partial x^-}$ and
${\brbeta=-2F\,\bar\partial x^+}$ recover the second-order action.
To absorb the explicit $L_\pm$ terms, one can redefine the multipliers as
${\beta'=\beta+2L_+\partial x^+}$ and
${\brbeta'=\brbeta+2L_-\bar\partial x^-}$.
At finite radius the remaining interaction is then
${(\beta'-2L_+\partial x^+)(\brbeta'-2L_-\bar\partial x^-)/(2F)}$;
the shifts do not remove its $L_\pm$ dependence.
This whole interaction vanishes toward the boundary, where
the first-order Gomis--Ooguri action is recovered.
On patches with ${F>0}$, its effective speed parameter is
\be
c_{\rm eff}^{\,2}:=2F\,,\qquad
\frac1{c_{\rm eff}^{\,2}}
=\frac{e^{-2y/l}}
 {2\left(1+L_+L_-e^{-4y/l}\right)}
\longrightarrow\frac12e^{-2y/l}\,.
 \label{SMceff}
\ee
Thus ${c_{\rm eff}\to\infty}$ follows from the radial asymptotics without
parameter tuning.  The $\beta\brbeta$ coupling restores relativistic
dynamics away from this limit.  For a constant coupling on the flat
type-$(1,1)$ background with isometries along $x^\pm$, this deformation
is related to TsT/$T\bar T$ deformations~\cite{Blair:2020ttbar}.
Here the coupling depends on the bulk fields, so the radial interpolation
alone does not define a $T\bar T$ flow.  Moreover, $c_{\rm eff}$ is not the
causal speed of the target-space metric: for constant ${L_\pm>0}$ it is
finite at the Killing horizon, with
${c_{\rm eff}^{\,2}=4\sqrt{L_+L_-}}$.
Let $g_\parallel$ denote the $2\times2$ metric induced on a constant-$y$
surface in the coordinates $(x^+,x^-)$, with components ${g_{++}=2L_+}$,
${g_{--}=2L_-}$ and ${g_{+-}=g_{-+}=-F}$.
If ${L_+L_-<0}$, an ${F=0}$ crossing makes the first-order parametrization
singular, while ${-\det g_\parallel=F^2-4L_+L_-}$ remains nonzero.

\subsection{Non-Riemannian Worldsheet and Marginal Deformations}
\label{SMsub:worldsheet:NR}

\subsubsection*{Clock Forms and the Exact Worldsheet Reduction}

The string Newton--Cartan data can be read directly from the
non-Riemannian generalized metric~\eqref{NRHcompact}
\cite{Andringa:2012uz,Harmark:2017rpg,Bergshoeff:2018yvt,
Bergshoeff:2019pij,Oling:2022fft}.  In the parametrization of
Ref.~\cite{Morand:2017fnv}, its two non-Riemannian covectors are
$X_\mu$ and $\bar X_\mu$, whose clock forms are
\be
 X=\cosh\tfrac{\chi}{2}\,\rd x^+
 -e^{-\sigma}\sinh\tfrac{\chi}{2}\,\rd x^-\,,
\qquad \bar X=\cosh\tfrac{\chi}{2}\,\rd x^-
 -e^{\sigma}\sinh\tfrac{\chi}{2}\,\rd x^+\,.
 \label{SNCtau}
\ee
Here $\chi$ and $\sigma$ are the target-space fields in~\eqref{NRHcompact};
the latter is distinct from the worldsheet coordinates $\sigma^\pm$.
Near the boundary,
${X=e^{-y/l}\vartheta^++\cO(e^{-4y/l})}$ and
${\bar X=e^{-y/l}\vartheta^-+\cO(e^{-4y/l})}$,
so their first corrections are ${-e^{-2y/l}L_-\rd x^-}$ and
${-e^{-2y/l}L_+\rd x^+}$, respectively.
The clock forms thus encode $L_\pm$ and match the rescaled Riemannian
vielbein at this order.  Their dual vectors and a Kalb--Ramond two-form
representative are
\be
 Y=\cosh\tfrac{\chi}{2}\,\partial_+
       +e^\sigma\sinh\tfrac{\chi}{2}\,\partial_-\,,
\quad \bar Y=e^{-\sigma}\sinh\tfrac{\chi}{2}\,\partial_+
       +\cosh\tfrac{\chi}{2}\,\partial_-\,,
\quad B_{(2)}=-\half\cW\,X\wedge\bar X
       =-\half\cW\,\rd x^+\wedge\rd x^-\,.
 \label{SNCdualB}
\ee
They obey ${Y\cdot X=\bar Y\cdot\bar X=1}$,
${Y\cdot\bar X=\bar Y\cdot X=0}$ and the completeness relation
${Y^\mu X_\nu+\bar Y^\mu\bar X_\nu+
\delta_y^\mu\delta_\nu^y=\delta^\mu{}_\nu}$.
Together with the radial tensors
${H^{\mu\nu}=\delta_y^\mu\delta_y^\nu}$ and
${K_{\mu\nu}=\delta_\mu^y\delta_\nu^y}$, these data give the mixed block
${\cH^\mu{}_\nu=Y^\mu X_\nu-\bar Y^\mu\bar X_\nu}$ and the lower-right block
\be
 \cH_{\mu\nu}
 =\delta_\mu^y\delta_\nu^y
 +2X_{(\mu}B_{\nu)\rho}Y^\rho
 -2\bar X_{(\mu}B_{\nu)\rho}\bar Y^\rho
 =\delta_\mu^y\delta_\nu^y
 +\cW(X_\mu\bar X_\nu+\bar X_\mu X_\nu)\,.
 \label{SNCreconstruction}
\ee
Write ${\mathfrak a_\mu=\mathfrak a_{+\mu}}$ and
${\bar{\mathfrak a}_\mu=\mathfrak a_{-\mu}}$, and define
${\beta=-Y^\mu\mathfrak a_\mu}$ and
${\brbeta=\bar Y^\mu\bar{\mathfrak a}_\mu}$.
The radial auxiliary equations give
${\mathfrak a_y=-\partial y}$ and
${\bar{\mathfrak a}_y=\bar\partial y}$.
Using completeness, their elimination yields the radial kinetic term,
while the two remaining multiplier combinations impose
\be
 X_{\mu}\bar\partial x^{\mu}=0\,,
 \qquad
 \bar X_{\mu}\partial x^{\mu}=0\,.
 \label{SMdygconstraints}
\ee
Before imposing these constraints, the lower-right block contributes
${\cW[(X\!\cdot\partial x)(\bar X\!\cdot\bar\partial x)
+(\bar X\!\cdot\partial x)(X\!\cdot\bar\partial x)]/(4\pi\alpha')}$.
The second product is absorbed by the unit-Jacobian redefinition
${\beta_{\rm new}=\beta_{\rm old}
+\half\cW\,\bar X_\mu\partial x^\mu}$.
Dropping the subscript on the new multiplier gives the exact Lagrangian
\be
 L_{\rm NR}=\frac{1}{2\pi\alpha'}\Big(
 \partial y\,\bar\partial y
 +\beta\,X_{\mu}\bar\partial x^{\mu}
 +\brbeta\,\bar X_{\mu}\partial x^{\mu}
 +\half\cW\,
 X_{\mu}\partial x^{\mu}\,
 \bar X_{\nu}\bar\partial x^{\nu}\Big)\,,
 \label{SMdygGO}
\ee
The additional target-space sectors and dilaton coupling are understood
separately.  In particular, no $\beta\brbeta$ term is generated: this branch has
${c_{\rm eff}^{-2}=0}$ at every radius.
At finite $\chi$, however, the clock forms are generally not closed and
cannot be replaced by coordinate differentials.  Equation~\eqref{SMdygGO}
therefore describes a curved first-order system.

As ${y\to\infty}$, ${\chi\to0}$, ${X\to\rd x^+}$ and ${\bar X\to\rd x^-}$.
The limiting kinetic terms are those of the
Gomis--Ooguri string~\cite{Gomis:2000bd,Lee:2013hma,Ko:2015rha},
\be
 L_{\rm GO}=\frac{1}{2\pi\alpha'}
 \big(\beta\,\bar\partial x^{+}+\brbeta\,\partial x^{-}
 +\partial y\,\bar\partial y\big)+\cdots\,,
 \label{SMGO}
\ee
where the ellipsis denotes the hair coupling and the additional sectors.
The multipliers have weights $(1,0)$ and $(0,1)$ and impose
${\bar\partial x^+=0=\partial x^-}$.
The limiting $x^\pm$ vertex inherited from~\eqref{SMdygGO} is
\be
 V_{\cW}=\frac{1}{4\pi\alpha'}\,
 \cW(x^{+},x^{-},y)\,\partial x^{+}\bar\partial x^{-}\,.
 \label{SMvertex}
\ee
The boundary value is ${\cW\to W_0}$; its subleading radial dressing is
analyzed below.  The free fields $(\beta,x^+)$, $(\brbeta,x^-)$ and $y$
also give the asymptotic Wakimoto description of the WZW long-string
sector~\cite{Giveon:1998ns,Knighton:2024,Knighton:2024b}.

\subsubsection*{The Boundary Theory and the $W_0$ Deformation}

We keep the multiplier normalization of~\eqref{SMGO}.
For points ${s=(s^+,s^-)}$ and ${u=(u^+,u^-)}$ on the Lorentzian worldsheet,
the nonzero free contractions are
\be
 \beta(s^+)x^+(u^+)\sim-\frac{\alpha'}{s^+-u^+}\,,\qquad 
 \brbeta(s^-)x^-(u^-)\sim-\frac{\alpha'}{s^--u^-}\,,
\qquad y(s)y(u)\sim-\frac{\alpha'}2
 \ln\!\big[(s^+-u^+)(s^--u^-)\big]\,.
\label{SMfreecontractions}
\ee
These are chiral short-distance expansions in real null separations;
the causal prescriptions are suppressed.  There are no $x^\pm$
self-contractions.  Varying the doubled action with respect to the
worldsheet metric and then using the auxiliary equations gives, at
${\cW=0}$ and before the dilaton improvement,
\be
 \begin{aligned}
 T&=\frac{4\pi}{\sqrt{-h}}\frac{\delta S_{\rm dyg}}{\delta h^{++}}
 =-\frac1{2\alpha'}\cH_{MN}D_+x^MD_+x^N
 =-\frac1{\alpha'}:\!\big[(\partial y)^2+\beta\partial x^+\big]\!:\,,\\
 \bar T&=\frac{4\pi}{\sqrt{-h}}\frac{\delta S_{\rm dyg}}{\delta h^{--}}
 =-\frac1{2\alpha'}\cH_{MN}D_-x^MD_-x^N
 =-\frac1{\alpha'}:\!\big[(\bar\partial y)^2+\brbeta\bar\partial x^-\big]\!:\,.
 \end{aligned}
\label{SMworldsheetstress}
\ee
Consequently, for any smooth $W_0(x^+,x^-)$,
\be
 T(s^+)V_{W_0}(u)\sim
 \frac{V_{W_0}(u)}{(s^+-u^+)^2}
 +\frac{\partial V_{W_0}(u)}{s^+-u^+}\,,
\qquad \bar T(s^-)V_{W_0}(u)\sim
 \frac{V_{W_0}(u)}{(s^--u^-)^2}
 +\frac{\bar\partial V_{W_0}(u)}{s^--u^-}\,.
\label{SMW0vertexOPE}
\ee
The vertex has weight $(1,1)$.  It contains neither multiplier nor a
radial dressing, so its self-OPE is nonsingular: $W_0$ is an exactly
marginal deformation of the boundary Gomis--Ooguri theory.

The Kalb--Ramond two-form description makes its local gauge character explicit.
The data in~\eqref{SNCdualB} satisfy
\be
 \cH(\cW)=\Omega_b\cH(0)\Omega_b^{T}\,,
 \qquad b_{+-}=-\half\cW\,.
 \label{SMWisB}
\ee
For ${\cW=W_0(x^+,x^-)}$, $b$ is closed and hence locally a Kalb--Ramond two-form
gauge transformation~\eqref{SMBtransform}.
In the ${E=g-B}$ convention, its worldsheet pullback contributes
\be
 \Delta L_{\rm ws}
 =\frac{1}{4\pi\alpha'}\,
 \cW\big(\partial x^{+}\bar\partial x^{-}
 -\bar\partial x^{+}\partial x^{-}\big)\,,
 \label{SMdeltaL}
\ee
Since ${X\wedge\bar X=\rd x^+\wedge\rd x^-}$, the difference
between this antisymmetric coupling and the symmetric coupling from
\eqref{SNCreconstruction} is proportional to the product of
\eqref{SMdygconstraints}.  A multiplier shift removes it, giving the
same vertex~\eqref{SMvertex}.  Local gauge triviality does not remove
$W_0$ from the boundary-source problem when the boundary data are fixed.
For the radially dependent mode, neither the regular self-OPE argument
nor the local gauge conclusion follows; both questions require the
radial sector.

\subsubsection*{Fradkin--Tseytlin Improvement and the Radial Hair}

On the common boundary vacuum, the radial action and the Lorentzian
generalized Fradkin--Tseytlin coupling are
\be
 S_{y}=\frac{1}{2\pi\alpha'}\int\rd^{2}\sigma\,
 \partial y\,\bar\partial y\,,
 \qquad
 S_{\rm FT}=-\frac{1}{4\pi}\int\rd^2\sigma\sqrt{-h}\,R^{(2)}d\,,
 \qquad d=-\frac{y}{l}\,.
 \label{SMFTsector}
\ee
The minus sign follows the Lorentzian convention of~\eqref{SMdyg}.
The curvature term must be varied before setting the worldsheet metric
to conformal gauge.  In two dimensions its variation contributes
${\Delta T_{\alpha\beta}=\trd_\alpha\trd_\beta d
-h_{\alpha\beta}\trd^2d}$, where $\trd_\alpha$ is the worldsheet
Levi-Civita derivative, and hence
${\Delta T=-\partial^2y/l}$ and
${\Delta\bar T=-\bar\partial^2y/l}$.
The flat-patch propagators are unchanged.  Since $W_0$ is independent
of $y$, this improvement leaves its preceding OPEs unchanged.

For comparison with the compact $S^3\times T^4$ completion, we fix
the level convention by the flux in~\eqref{SMuplift}:
\be
 k\equiv N_{5}
 =\frac{1}{4\pi^{2}\alpha'}\left|\int_{S^{3}}H\right|
 =\frac{l^{2}}{\alpha'}\in\mathbb Z_{>0}\,,
 \qquad
 q_{\rm ws}\equiv\frac{\alpha'}{l^{2}}=\frac1{k}\,.
 \label{SMWZWlevel}
\ee
Here $k$ is the supersymmetric WZW level.  After decoupling the adjoint
fermions, the bosonic $\SL(2,\mathbb R)$ level is
${k_{\rm bos}=k+2=2+l^2/\alpha'}$, while the bosonic $\SU(2)$ level
is $k-2$~\cite{Maldacena:2000hw,Giribet:2018ada,Eberhardt:2018ouy}.

With the dilaton improvement included, the radial stress tensor and
exponential weight are
\be
 T_{y}=-\frac1{\alpha'}:\!(\partial y)^{2}\!:
      -\frac1l\partial^{2}y\,,
 \qquad
 h_{y}(a)=-\frac{\alpha'}4a\left(a+\frac2l\right)\,,
\qquad T_{y}(s^+)e^{ay(u)}
 \sim\frac{h_{y}(a)}{(s^+-u^+)^{2}}e^{ay(u)}
 +\frac{\partial e^{ay(u)}}{s^+-u^+}\,.
 \label{SMFTweight}
\ee
The full left-moving stress tensor is
${T=T_y-\frac1{\alpha'}:\!\beta\partial x^+\!:}$, with an identical
right-moving construction.  Thus a vertex
${\mathcal U_f=f(x^+,x^-,y)\partial x^+\bar\partial x^-}$ has
weight $(1,1)$ precisely when the additional radial double pole vanishes:
\be
 \left(\partial_{y}^{2}+\frac2l\partial_{y}\right)f=0\,,
 \qquad
 f=W_{0}+e^{-2y/l}W_{1}\,.
 \label{SMBRSTradial}
\ee
Since arbitrary
$W_1(x^+,x^-)$ has weight $(0,0)$ in the $(\beta,x^+)$ and
$(\brbeta,x^-)$ first-order sectors,
\be
 \mathcal U_{W_{1}}
 =e^{-2y/l}W_{1}(x^{+},x^{-})
   \partial x^{+}\bar\partial x^{-}
 \quad\hbox{is a matter primary of weight }(1,1)\,.
 \label{SMBRSTvertex}
\ee
Its action coefficient is $1/(4\pi\alpha')$, as in~\eqref{SMvertex}.
This establishes the linear marginality of the asymptotic hair
deformation and continues the lower-right generalized-metric
polarization of the constant-background spectrum~\cite{Park:2020ixf}.
It is a bosonic matter Virasoro condition; a type-II super-BRST and
GSO analysis is not supplied here.
Exact marginality at finite $W_1$ is not established here.

\subsubsection*{Central Charge and the Scope of the Vertex Analysis}

The fourth-order pole in the $TT$ OPE is
${(3+6\alpha'/l^2)/[2(s^+-u^+)^4]}$.
The commuting $(\beta,x^+)$ pair contributes two units of central
charge, and the improved radial field contributes
\be
 c_{y}=1+\frac{6\alpha'}{l^{2}}\,,
 \qquad
 c_{\beta\gamma}+c_y
 =2+\left(1+\frac6{k}\right)
 =\frac{3(k+2)}{k}\,,
 \label{SMBRSTcentral}
\ee
Equivalently, this is ${3k_{\rm bos}/(k_{\rm bos}-2)}$, the central
charge of the bosonic $\SL(2,\mathbb R)_{k_{\rm bos}}$ WZW model.
The right-moving sector has the same central charge.

Around ${L_\pm=W_0=0}$, consider a lower-right perturbation with only
${\delta\cH_{+-}=\delta\cH_{-+}=e^{-2y/l}W_1}$ nonzero.
For the generalized Lie derivative along
${\xi^M=(\tilde{\lambda}_\mu,v^\mu)}$, the conditions
${\hcL_\xi\cH_{MN}=\delta\cH_{MN}}$ and ${\hcL_\xi d=0}$ require
${b_{+-}=-\half e^{-2y/l}W_1}$, ${b_{+y}=\partial_+v^y}$,
${b_{-y}=-\partial_-v^y}$ and ${\partial_+\partial_-v^y=0}$,
where ${b=\rd\tilde{\lambda}}$.  Hence
\be
 0=(\rd b)_{+-y}
 =-\half\partial_y\big(e^{-2y/l}W_1\big)
 =\frac1l e^{-2y/l}W_1\,.
 \label{SMWgaugeobstruction}
\ee
Thus a nonzero $W_1$ is not pure generalized gauge around this vacuum.

\section{Ten-Dimensional Uplift and Killing Spinors}
\label{SMsec:susy}

\subsection{Ten-Dimensional Killing-Spinor Equations}
\label{SMsub:susy:equations}

Both the Riemannian and the non-Riemannian three-dimensional backgrounds can
be embedded into ten dimensions by adjoining a three-sphere supported by
Kalb--Ramond flux and a flat four-dimensional space. We parametrize the sphere
by $(\theta,\varphi_1,\varphi_2)$ with ${0<\theta<\pi/2}$ and the flat directions
by $(z^1,z^2,z^3,z^4)$, and we take
\be
\resizebox{0.85\textwidth}{!}{$\displaystyle
\rd s^2_{S^3}=l^2\big(\rd\theta^2+\cos^2\theta\,\rd\varphi_1^2
+\sin^2\theta\,\rd\varphi_2^2\big)\,,\quad
B_{S^3}=l^2\sin^2\theta\,\rd\varphi_1\wedge\rd\varphi_2\,,\quad
\rd s^2_{\mathbb R^4}=\delta_{mn}\rd z^m\rd z^n\,,\quad B_{\mathbb R^4}=0\,.
$}
\label{SMuplift}
\ee
The flat factor may be compactified to $T^4$ with periodic spin structure.
The DFT variables are uplifted as
\be
\cH_{10}=\cH_3\oplus\cH_{S^3}\oplus\cH_{\mathbb R^4}\,,\qquad
d_{10}=d_3-\half\ln(l^3\sin\theta\cos\theta)\,.
\label{SMupliftDFT}
\ee
With the sphere flux, both uplifts satisfy ${G^{(10)}_{\hat M\hat N}=0}$.
For the Riemannian three-dimensional block we use
$(\vartheta^+,\vartheta^-,\rd y)$ in both frames; for the non-Riemannian block
we use~\eqref{SMvielbein}. The source $W_0$ drops out of the local
Killing-spinor equations. Both internal coframes are
$(l\rd\theta,l\cos\theta\,\rd\varphi_1,l\sin\theta\,\rd\varphi_2,\rd z^m)$.

Type-II supersymmetric DFT contains two Majorana--Weyl spinors $\varepsilon$ and $\varepsilon'$,
which transform under the two local Lorentz groups. On a purely bosonic
NS--NS background, the vanishing of the supersymmetry variations of the
gravitini and dilatini reduces to the Killing-spinor equations~\cite{Jeon:2012hp}
\be
\cD_{\hat{\bar p}}\varepsilon=0\,,\qquad
\gamma^{\hat p}\cD_{\hat p}\varepsilon=0\,,\qquad
\cD_{\hat p}\varepsilon'=0\,,\qquad
\bar\gamma^{\hat{\bar p}}\cD_{\hat{\bar p}}\varepsilon'=0\,.
\label{SMtypeIIKS}
\ee
The hatted indices are ten-dimensional, and
${\cD_{\hat p}=V^{\hat M}{}_{\hat p}\cD_{\hat M}}$ and
${\cD_{\hat{\bar p}}=\brV^{\hat M}{}_{\hat{\bar p}}\cD_{\hat M}}$
are the local Lorentz projections of the master derivative, which acts as
\be
\begin{aligned}
\cD_{\hat M}\varepsilon
&=\partial_{\hat M}\varepsilon+\quarter\Phi_{\hat M\hat p\hat q}
\gamma^{\hat p\hat q}\varepsilon\,,\qquad
\Phi_{\hat M\hat p\hat q}=V^{\hat N}{}_{\hat p}\nabla_{\hat M}V_{\hat N\hat q}\,,\\
\cD_{\hat M}\varepsilon'
&=\partial_{\hat M}\varepsilon'+\quarter\bar\Phi_{\hat M\hat{\bar p}\hat{\bar q}}
\bar\gamma^{\hat{\bar p}\hat{\bar q}}\varepsilon'\,,\qquad
\bar\Phi_{\hat M\hat{\bar p}\hat{\bar q}}
=\brV^{\hat N}{}_{\hat{\bar p}}\nabla_{\hat M}\brV_{\hat N\hat{\bar q}}\,.
\end{aligned}
\label{SMspinorderivatives}
\ee
Here ${\nabla_{\hat M}V_{\hat N\hat q}=\partial_{\hat M}V_{\hat N\hat q}
+\Gamma_{\hat M\hat N}{}^{\hat P}V_{\hat P\hat q}}$, with the DFT Christoffel
connection~\eqref{SMbackgroundconnection} evaluated on the full ten-dimensional
background, ${D=10}$; fermionic torsions vanish. The first and third equations in~\eqref{SMtypeIIKS} are the
gravitino equations, and the second and fourth are the dilatino equations.
The spinor $\varepsilon$ is differentiated along the barred frame but rotated
by the unbarred spin connection, and vice versa for $\varepsilon'$.

We realize the ten-dimensional Clifford algebra
${\{\gamma^{\hat p},\gamma^{\hat q}\}=2\eta^{\hat p\hat q}}$ on
$\mathbb C^2\otimes\mathbb C^2\otimes\mathbb C^2\otimes\mathbb C^4$ as
\be
\begin{gathered}
\gamma^{\hat p}=\lambda^p\otimes I_2\otimes\sigma_1\otimes I_4\,,\qquad
\gamma^{\hat{\imath}}=I_2\otimes\sigma_i\otimes\sigma_2\otimes I_4\,,\qquad
\gamma^{\hat m}=I_2\otimes I_2\otimes\sigma_3\otimes\rho_m\,,\\
\lambda^\oplus=\begin{pmatrix}0&0\\\sqrt2&0\end{pmatrix}\,,\qquad
\lambda^\ominus=\begin{pmatrix}0&-\sqrt2\\0&0\end{pmatrix}\,,\qquad
\lambda^{\mathrm y}=\sigma_3\,.
\end{gathered}
\label{SMgammaten}
\ee
Here ${p=(\oplus,\ominus,\mathrm y)}$ labels the three-dimensional subset
of $\hat p$ in the first expression, ${i=1,2,3}$ labels
$(\theta,\varphi_1,\varphi_2)$, and ${m=1,\ldots,4}$.
These are $\Spin(1,9)$ vector indices; their $\Spin(9,1)$ counterparts
are $(\brq,\bar{\imath},\bar m)$. The $\sigma_i$ are Pauli matrices.
The Euclidean matrices are
${\rho_{1,2}=\sigma_{1,2}\otimes I_2}$,
${\rho_{3,4}=\sigma_3\otimes\sigma_{1,2}}$, and
${\rho_5=\sigma_3\otimes\sigma_3}$.
We use ${\eta_{\oplus\ominus}=-1}$, ${\eta_{\mathrm{yy}}=1}$, and ${\bar\eta=-\eta}$.
The chirality matrix and the Majorana intertwiner are
\be
\begin{gathered}
\gamma^{(11)}=I_2\otimes I_2\otimes\sigma_3\otimes\rho_5\,,\qquad
\mathsf B_4=\sigma_1\otimes\sigma_2\,,\qquad
\mathsf B_{10}=I_2\otimes\sigma_2\otimes I_2\otimes\mathsf B_4\,,\\
\mathsf B_{10}\gamma^{\hat p}\mathsf B_{10}^{-1}=(\gamma^{\hat p})^*\,,\qquad
\bar\gamma^{\hat{\bar p}}=\gamma^{(11)}\gamma^{\hat p}\,,\qquad
\bar\gamma^{(11)}=-\gamma^{(11)}\,,\qquad
\bar\gamma^{\hat{\bar p}\hat{\bar q}}=-\gamma^{\hat p\hat q}\,.
\end{gathered}
\label{SMgammaMajorana}
\ee
The barred matrices obey
${\{\bar\gamma^{\hat{\bar p}},\bar\gamma^{\hat{\bar q}}\}=2\bar\eta^{\hat{\bar p}\hat{\bar q}}}$
and have the same Majorana intertwiner. We impose
\be
\gamma^{(11)}\varepsilon=\varepsilon\,,\qquad
\mathsf B_{10}\varepsilon=\varepsilon^*\,,\qquad
\bar\gamma^{(11)}\varepsilon'=-\varepsilon'\,,\qquad
\mathsf B_{10}\varepsilon'=\varepsilon'{}^*\,.
\label{SMspinorreality}
\ee
Each spinor has sixteen real components. The four factors respectively
describe the three-dimensional spinor, the sphere spinor, an auxiliary
two-dimensional factor, and the flat internal spinor. Accordingly, we write
\be
\Psi_{\alpha\beta\gamma\delta}\,,\qquad
\Psi=\varepsilon,\varepsilon'\,,\qquad
\alpha,\beta,\gamma=1,2\,,\qquad\delta=1,\ldots,4\,.
\label{SMfourspinorindices}
\ee

\subsection{Killing Spinors under the Periodic Identification}
\label{SMsub:susy:periodic}

We impose ${\varphi\sim\varphi+2\pi}$, which acts as
${x^\pm\mapsto x^\pm\pm\sqrt2\pi l}$, with periodic boundary conditions for both
spinors. The Killing-spinor equations are first solved locally and periodicity
is imposed afterwards. For the Riemannian branch we take $L_\pm$ to be zero or
positive constants. The one-sided non-Riemannian families below also allow
arbitrary smooth profiles of the surviving $L_\pm$.
We describe the solutions by their nonvanishing components, their
$x^\pm$ dependence, and their number of real parameters. Their sphere dependence
is fixed by the Killing-spinor equations; the Riemannian periodic solutions
also carry the radial factor $e^{y/(2l)}$.
We write ${\cN=(n_+,n_-)}$ for the numbers of real $x^\pm$-independent Killing
spinors in the $\varepsilon$ and $\varepsilon'$ sectors, respectively.
These count constant representatives, not the full function space or
fermionic phase-space charges. In all component lists below, ${\beta=1,2}$
and the Majorana condition is understood.

\paragraph{Two-sided positive constants, ${L_+>0}$ and ${L_->0}$.}
Neither branch admits periodic Killing spinors. In the Riemannian branch the
local solutions depend on $x^\pm$ through
$e^{\pm\sqrt{2L_\pm}\,x^\pm/l}$, so that no solution is periodic~\cite{Coussaert:1993}.
In the non-Riemannian branch there are no local solutions: the algebraic
gravitino constraint and its differential consistency force both
three-dimensional spinor components to vanish, independently of $W_1$.
In both branches ${\cN=(0,0)}$.

\paragraph{One-sided profiles, ${L_-=0}$ with ${L_+\neq0}$.}
Here the two branches have $x^\pm$-independent Killing spinors in opposite
sectors when ${W_1\neq0}$. For constant ${L_+>0}$, the Riemannian branch
has ${\varepsilon=0}$, while $\varepsilon'$ has the nonvanishing components
$\varepsilon'_{2\beta2\delta}$, ${\delta=2,3}$, which are independent of
$x^\pm$ and carry four real parameters, so that ${\cN=(0,4)}$.
In the non-Riemannian branch with generic nonzero $W_1$, ${\varepsilon'=0}$,
while $\varepsilon$ has the nonvanishing components
$\varepsilon_{2\beta1\delta}$, ${\delta=1,4}$, parametrized by four arbitrary
real periodic functions of $x^+$. Thus ${\cN=(4,0)}$, with an infinite tower of
nonconstant modes. This family does not depend on the form of
$L_+(x^+)$ or $W_1(x^+,x^-)$.
With ${W_1=0}$, both families are present: the four arbitrary periodic
functions in $\varepsilon_{2\beta1\delta}$, ${\delta=1,4}$, and four
$x^\pm$-independent parameters in $\varepsilon'_{2\beta2\delta}$,
${\delta=2,3}$, giving ${\cN=(4,4)}$.
The added constant modes have the Riemannian polarizations, up to the radial
factor. Turning on $W_1$ removes these constant modes and leaves the
arbitrary periodic profiles.

\paragraph{One-sided profiles, ${L_+=0}$ with ${L_-\neq0}$.}
The mirror statements hold with ${\varepsilon\leftrightarrow\varepsilon'}$
and ${x^+\leftrightarrow x^-}$.
For constant ${L_->0}$, the Riemannian branch has ${\varepsilon'=0}$,
while $\varepsilon$ has the nonvanishing components $\varepsilon_{1\beta2\delta}$, ${\delta=2,3}$,
independent of $x^\pm$ and carrying four real parameters, so that
${\cN=(4,0)}$.
In the non-Riemannian branch with generic nonzero $W_1$, ${\varepsilon=0}$,
while $\varepsilon'$ has the nonvanishing components
$\varepsilon'_{1\beta1\delta}$, ${\delta=1,4}$, parametrized by four arbitrary
real periodic functions of $x^-$, so that ${\cN=(0,4)}$.
With ${W_1=0}$, both families are present and ${\cN=(4,4)}$.

Special nonzero hair profiles can admit additional nonconstant solutions
in the opposite sector. On ${L_-=0}$, the profiles in
$(\varepsilon'_{1\beta2\delta},\varepsilon'_{2\beta2\delta})$
are $((l/\sqrt2)\partial_-f_-^r,f_-^r)$; on ${L_+=0}$, those in
$(\varepsilon_{1\beta2\delta},\varepsilon_{2\beta2\delta})$
are $(f_+^r,(l/\sqrt2)\partial_+f_+^r)$, with ${\delta=2,3}$.
For each of the four real internal polarizations, the periodic functions satisfy
\be
\begin{aligned}
L_-=0:&\qquad 2l^2L_+(x^+)\partial_-^2f_-^r(x^-)
=W_1(x^+,x^-)f_-^r(x^-)\,,\\
L_+=0:&\qquad 2l^2L_-(x^-)\partial_+^2f_+^r(x^+)
=W_1(x^+,x^-)f_+^r(x^+)\,,\qquad r=1,\ldots,4\,.
\end{aligned}
\label{SMhaircompatibility}
\ee
For example, ${W_1=-4n^2L_+}$ on ${L_-=0}$ admits
${f_-^r\propto\cos(\sqrt2 n x^-/l)}$ or $\sin(\sqrt2 n x^-/l)$ for integer
${n\neq0}$. These extra modes do not alter the $x^\pm$-independent counts.

\paragraph{The case ${L_+=L_-=0}$ with ${W_1\neq0}$.}
Both sectors are now populated in each branch, and their
$x^\pm$-independent spinors occupy complementary sets of components.
In the Riemannian branch $\varepsilon$ has the nonvanishing components
$\varepsilon_{1\beta2\delta}$ and $\varepsilon'$ has
$\varepsilon'_{2\beta2\delta}$, with ${\delta=2,3}$ in both cases.
They are independent of $x^\pm$ and carry four real parameters each,
so that ${\cN=(4,4)}$.
For arbitrary $W_1$, the non-Riemannian families have
$\varepsilon_{2\beta1\delta}$ and $\varepsilon'_{1\beta1\delta}$,
with ${\delta=1,4}$, parametrized by four arbitrary real periodic functions
of $x^+$ and $x^-$, respectively. Again ${\cN=(4,4)}$;
additional nonconstant modes, if present, obey~\eqref{SMhaircompatibility}.

\paragraph{The non-Riemannian vacuum, ${L_+=L_-=W_1=0}$.}
The spinor $\varepsilon$ now carries eight arbitrary real periodic functions
of $x^+$: four sit in $\varepsilon_{2\beta1\delta}$, ${\delta=1,4}$, as before,
and the other four sit in $\varepsilon_{1\beta2\delta}$, ${\delta=2,3}$,
and are accompanied by their $x^+$-derivatives in
$\varepsilon_{2\beta2\delta}$, ${\delta=2,3}$.
The derivative components are nonvanishing but not independent.
Likewise $\varepsilon'$ carries eight arbitrary real periodic functions
of $x^-$: four in $\varepsilon'_{1\beta1\delta}$, ${\delta=1,4}$, and four
in $\varepsilon'_{2\beta2\delta}$, ${\delta=2,3}$, the latter accompanied
by their $x^-$-derivatives in $\varepsilon'_{1\beta2\delta}$, ${\delta=2,3}$.
For $x^\pm$-independent Killing spinors the derivative components vanish,
and one finds ${\cN=(8,8)}$, the union of the Riemannian and the hairy
non-Riemannian component sets above.
The Riemannian periodic spinors are recovered, up to their radial factors,
by keeping the $x^\pm$-independent parts of the former set, while
turning on $W_1$ retains only the latter set of constant representatives.

In all cases with ${L_+L_-=0}$ the non-Riemannian branch is therefore
supersymmetric for arbitrary smooth profiles of the surviving $L_\pm$ and
$W_1(x^+,x^-)$, and its Killing spinors form an infinite-dimensional space,
parametrized by periodic functions with modes
$e^{i\sqrt2 n x^\pm/l}$, ${n\in\mathbb Z}$.
At ${L_+=L_-=0}$, the $x^\pm$-independent Riemannian and hairy
non-Riemannian representatives are complementary halves of the sixteen
real constant Killing spinors of the non-Riemannian vacuum.

\section{A Free Abelian CFT on the DFT Boundary}
\label{SMsec:candidate}
\begingroup
\setlength{\abovedisplayskip}{3pt plus 1pt minus 1pt}
\setlength{\belowdisplayskip}{3pt plus 1pt minus 1pt}
\setlength{\abovedisplayshortskip}{2pt plus 1pt minus 1pt}
\setlength{\belowdisplayshortskip}{2pt plus 1pt minus 1pt}
\setlength{\jot}{6pt}

The constant non-Riemannian boundary DFT vielbeins~\eqref{Rboundaryframe}
support a free abelian model whose propagating sector is a CFT.
Its fields are a real bispinor $\cB^\alpha{}_{\bar\alpha}$
and Grassmann-odd spinors $\psi^\alpha,\psi'{}^{\bar\alpha}$ under the local
Lorentz groups $\Spin(1,1)\times\Spin(1,1)$.
All fields have diffeomorphism weight $\half$.  No chirality projection
on $\cB$ or self-duality condition on its field strength is imposed.
We use the Ramond--Ramond (RR) differential complex of~\cite{Jeon:2012rr}.

The gamma matrices obey the Clifford algebras of $\eta_{ab}$ and
$\bar\eta_{\bar a\bar b}$ in~\eqref{Rboundaryframe}.
The chirality matrices satisfy ${(\gamma^{(3)})^2=(\bar\gamma^{(3)})^2=1}$;
define ${\gamma^{ab}:=\half[\gamma^a,\gamma^b]}$ and its barred counterpart.
Choose the symmetric charge-conjugation matrices ${C=C_+}$ and
${\bar C=\bar C_+}$ of~\cite{Jeon:2012rr}:
\be
 \begin{gathered}
 C^T=C\,,\qquad (\gamma^a)^T=C\gamma^aC^{-1}\,,\qquad
 \bar C^T=\bar C\,,\qquad
 (\bar\gamma^{\bar a})^T=\bar C\bar\gamma^{\bar a}\bar C^{-1}\,,\\
 \bar\psi:=\psi^TC\,,\qquad
 \bar\psi':=\psi'{}^T\bar C\,,\qquad
 \overline{\cB}:=\bar C^{-1}\cB^T C\,.
 \end{gathered}
 \label{SMcandidateconjugation}
\ee
The same conjugation applies to $\cF$.

The torsionless master derivative ${\cD_A=\nabla_A+\Phi_A+\bar\Phi_A}$,
with ${\nabla_A=\partial_A+\Gamma_A}$, acts on the weight-$\half$
fields as~\cite{Jeon:2012rr,Park:2025core}
\be
 \begin{aligned}
 \cD_A \cB
 &=\partial_A \cB-\half\Gamma^B{}_{BA}\cB+\Phi_A \cB-\cB\bar\Phi_A\,,
 &\Phi_A&:=\quarter\Phi_{Aab}\gamma^{ab}\,,
 &\bar\Phi_A&:=\quarter\bar\Phi_{A\bar a\bar b}
                                      \bar\gamma^{\bar a\bar b}\,,\\
 \cD_A\psi
 &=(\partial_A-\half\Gamma^B{}_{BA}+\Phi_A)\psi\,,
 &\cD_A\psi'
 &=(\partial_A-\half\Gamma^B{}_{BA}+\bar\Phi_A)\psi'\,.
 \end{aligned}
 \label{SMcandidateD}
\ee
Set ${\cD_a:=V^A{}_a\cD_A}$ and
${\cD_{\bar a}:=\brV^A{}_{\bar a}\cD_A}$; similarly,
${\gamma^A:=V^A{}_a\gamma^a}$ and
${\bar\gamma^A:=\brV^A{}_{\bar a}\bar\gamma^{\bar a}}$.  Define
\be
 \boxed{\cD_\pm \cB:=\gamma^a\cD_a \cB
              \pm\gamma^{(3)}(\cD_{\bar a}\cB)\bar\gamma^{\bar a}\,,\qquad
        \cD_\pm^{\,2}\cB=0\,,\qquad \cF:=\cD_+\cB\,.}
 \label{SMcandidateRRcomplex}
\ee
Nilpotency holds on weight-$\half$ bispinors subject to the section
condition~\cite{Jeon:2012rr}.  The density term in~\eqref{SMcandidateD},
with ${\Gamma^B{}_{BA}=-2\partial_A d}$, cancels the $d$-dependent contracted
spin connections.  Thus $\cD_\pm$ and both Dirac operators are independent
of $d$ at fixed DFT vielbeins.

The action is
\be
 \boxed{
 S_\partial=\int\rd^2x\,\cL_\partial\,,\qquad
 \cL_\partial=-\half\Tr(\cF\overline{\cF})
       +\bar\psi\gamma^a\cD_a\psi
       +\bar\psi'\bar\gamma^{\bar a}\cD_{\bar a}\psi'\,.}
 \label{SMcandidateaction}
\ee
The integrand has density weight one and requires no $e^{-2d}$ measure.
The dilaton decouples: ${\delta S_\partial/\delta d=0}$ at fixed fields
and DFT vielbeins.  The equations of motion are
\be
 \cD_-\cF=0\,,\qquad
 \gamma^a\cD_a\psi=0\,,\qquad
 \bar\gamma^{\bar a}\cD_{\bar a}\psi'=0\,.
 \label{SMcandidateEOM}
\ee

\paragraph{Kosmann transformations and Witt symmetry.}
For spinorial densities of weight $\omega$, the generalized Kosmann
derivative~\cite{Park:2025core} is
\be
 \begin{gathered}
 \fcL_\xi\cB
 =\xi^A\cD_A\cB+\omega(\cD_A\xi^A)\cB
   +\half\cD_{[a}\xi_{b]}\gamma^{ab}\cB
   -\half\cD_{[\bar a}\xi_{\bar b]}\cB\bar\gamma^{\bar a\bar b}\,,\\
 \fcL_\xi\psi
 =\xi^A\cD_A\psi+\omega(\cD_A\xi^A)\psi
   +\half\cD_{[a}\xi_{b]}\gamma^{ab}\psi\,,\quad
 \fcL_\xi\psi'
 =\xi^A\cD_A\psi'+\omega(\cD_A\xi^A)\psi'
   +\half\cD_{[\bar a}\xi_{\bar b]}
                                \bar\gamma^{\bar a\bar b}\psi'\,.
 \end{gathered}
 \label{SMcandidateKosmann}
\ee
Here ${\omega=\half}$ for $\cB,\psi,\psi'$ and ${\omega=0}$ for
$\varepsilon,\varepsilon'$.
When the DFT vielbeins also transform, the action is covariant under
generalized diffeomorphisms and both local Lorentz groups.  On the fixed background $V^{(0)}_{Aa},\brV^{(0)}_{A\bra}$
of~\eqref{Rboundaryframe}, take
${\xi^A=(0;v^\mu)}$ with ${v^\pm=v^\pm(x^\pm)}$, for which
${\fcL_\xi\cH^{(0)}_{AB}=\hcL_\xi\cH^{(0)}_{AB}=0}$ and
${\fcL_\xi V^{(0)}_{Aa}=\fcL_\xi\brV^{(0)}_{A\bra}=0}$.  Then
\be
 \begin{gathered}
 \delta_v\cB=\fcL_v\cB\,,\qquad
 \delta_v\psi=\fcL_v\psi\,,\qquad
 \delta_v\psi'=\fcL_v\psi'\,,\qquad
 \delta_v\cL_\partial=\partial_\mu(v^\mu\cL_\partial)\,,\\
 [\fcL_v,\fcL_w]=\fcL_{[v,w]}\,,\qquad
 [v,w]^\pm=v^\pm\partial_\pm w^\pm-w^\pm\partial_\pm v^\pm\,.
 \end{gathered}
 \label{SMcandidateWitt}
\ee
These are two commuting Witt algebras.  Since $d$ decouples, no condition
${\hcL_\xi d=0}$ is imposed on these generalized-metric isometries.

\paragraph{Abelian RR gauge symmetry.}
A weight-$\half$ bispinor $\lambda$ generates the RR gauge symmetry:
\be
 \delta_\lambda\cB=\cD_+\lambda\,,\qquad
 \delta_\lambda\cF=0\,,\qquad
 \delta_\lambda\psi=\delta_\lambda\psi'=0\,,
 \qquad \lambda\sim\lambda+\cD_+\lambda_1\,.
 \label{SMcandidategauge}
\ee
Both invariance and reducibility follow from ${\cD_+^2=0}$.

\paragraph{Supersupersymmetry.}
Choose weight-zero Grassmann-odd two-component parameters
$\varepsilon^\alpha,\varepsilon'{}^{\bar\alpha}$ satisfying the covariant
conformal Killing-spinor conditions on~\eqref{Rboundaryframe}:
\be
 \begin{aligned}
 (1+\gamma^{(3)})\varepsilon&=0\,,
 & (1+\bar\gamma^{(3)})\varepsilon'&=0\,,\\
 \cD_{\bar a}\varepsilon&=0\,,
 & \cD_a\varepsilon'&=0\,,\\
 \cD_a\varepsilon&=\half\gamma_a\gamma^b\cD_b\varepsilon\,,
 & \cD_{\bar a}\varepsilon'
   &=\half\bar\gamma_{\bar a}\bar\gamma^{\bar b}
                                        \cD_{\bar b}\varepsilon'\,.
 \end{aligned}
 \label{SMcandidateparameters}
\ee
The first row selects the supersymmetries acting on the propagating
fields; opposite parameter chiralities act only on fields absent from
the fixed-background action.  The last row gives the two-dimensional twistor equations.
Arbitrary chiral solutions generate supersupersymmetry:
\be
 \boxed{\delta_\varepsilon\cB
       =\psi\bar\varepsilon'-\varepsilon\bar\psi'\,,\qquad
 \delta_\varepsilon\psi=\gamma^{(3)}\cF\varepsilon'\,,\qquad
 \delta_\varepsilon\psi'=\overline{\cF}\varepsilon\,.}
 \label{SMcandidatefermionic}
\ee
Off shell,
\be
 \delta_\varepsilon\cL_\partial
 =\partial_A\!\left(
   -\bar\psi'\overline{\cF}\gamma^A\varepsilon
   +\bar\psi\gamma^{(3)}\cF\bar\gamma^A\varepsilon'\right)\,.
 \label{SMcandidatefermionicL}
\ee
For a chiral vector $v$, define the Cartan operator
\be
 \iota_v\cB:=\half v^A(\gamma_A \cB+\gamma^{(3)} \cB\bar\gamma_A)\,,
 \qquad
 \{\cD_+,\iota_v\}=\fcL_v\,.
 \label{SMcandidateCartan}
\ee
The fermionic transformations close as
\be
 \begin{gathered}
 \relax[\delta_1,\delta_2]\cB
 =\iota_v\cF=\fcL_v\cB-\cD_+(\iota_v\cB)\,,\\
 [\delta_1,\delta_2]\psi\doteq\fcL_v\psi\,,\qquad
 [\delta_1,\delta_2]\psi'\doteq\fcL_v\psi'\,,\\
 v^A=2\bar\varepsilon_1\gamma^A\varepsilon_2
       +2\bar\varepsilon'_1\bar\gamma^A\varepsilon'_2\,.
 \end{gathered}
 \label{SMcandidateclosure}
\ee
Here $\doteq$ allows fermion equations of motion and shifts of fields
absent from the fixed-background action.  The propagating sector realizes classical
${\mathcal N=(1,1)}$ superconformal symmetry.

\paragraph{Degenerate and equation-of-motion symmetries.}
On this fixed background, fields absent from the action admit arbitrary
shifts, not all of which are RR gauge transformations.  There are also trivial fermionic symmetries
${\delta\Psi=M(x)E}$, where ${\Psi=(\psi,\psi')^T}$,
$\delta S_\partial=\int\rd^2x\,\delta\Psi^T E$ defines the left Euler
derivative, and ${M=M^T}$ is even.  Grassmann antisymmetry gives ${E^TME=0}$;
these variations vanish on shell and carry no independent on-shell charges.

\paragraph{Explicit gamma matrices and component reduction.}

With the constant DFT vielbeins~\eqref{Rboundaryframe} and constant $d$,
all connections vanish, ${\Gamma_{ABC}=\Phi_{Aab}=\bar\Phi_{A\bar a\bar b}=0}$,
and ${\cD_A=\partial_A}$.

In the ordered null bases $(\oplus,\ominus)$ and
$(\bar\oplus,\bar\ominus)$, take ${\gamma^\oplus=\lambda^\oplus}$ and
${\gamma^\ominus=\lambda^\ominus}$ from~\eqref{SMgammaten}:
\be
 \begin{gathered}
 \eta=\begin{pmatrix}0&-1\\-1&0\end{pmatrix}\,,\qquad
 \bar\eta=-\eta\,,\qquad
 \gamma^\oplus=\sqrt2\begin{pmatrix}0&0\\1&0\end{pmatrix}\,,\qquad
 \gamma^\ominus=-\sqrt2\begin{pmatrix}0&1\\0&0\end{pmatrix}\,,\\
 \bar\gamma^{\bar\oplus}=\gamma^\oplus\,,\qquad
 \bar\gamma^{\bar\ominus}=-\gamma^\ominus\,,\qquad
 C=\bar C=\begin{pmatrix}0&1\\1&0\end{pmatrix}\,,\qquad
 \gamma^{(3)}=\begin{pmatrix}-1&0\\0&1\end{pmatrix}\,,\qquad
 \bar\gamma^{(3)}=-\gamma^{(3)}\,.
 \end{gathered}
 \label{SMcandidategammas}
\ee
Here ${\gamma^{(3)}=\gamma^0\gamma^1}$ with
${\gamma^0=(\gamma^\oplus+\gamma^\ominus)/\sqrt2}$ and
${\gamma^1=(\gamma^\oplus-\gamma^\ominus)/\sqrt2}$; the same definition
with barred matrices gives $\bar\gamma^{(3)}$.
The differential operators reduce to
\be
 \begin{gathered}
 \cD_\pm \cB=\frac{1}{\sqrt2}\bigl(
 \gamma^\oplus\partial_+\cB
 \pm\gamma^{(3)}(\partial_-\cB)\bar\gamma^{\bar\ominus}\bigr)\,,\\
 \gamma^a\cD_a\psi=\frac{1}{\sqrt2}\gamma^\oplus\partial_+\psi\,,
 \qquad
 \bar\gamma^{\bar a}\cD_{\bar a}\psi'=\frac{1}{\sqrt2}
                    \bar\gamma^{\bar\ominus}\partial_-\psi'\,.
 \end{gathered}
 \label{SMcandidateflatD}
\ee
Nilpotency follows from
${(\gamma^\oplus)^2=(\bar\gamma^{\bar\ominus})^2=0}$ and
${\{\gamma^\oplus,\gamma^{(3)}\}=0}$.

Write the unrestricted fields as
\be
 \cB=\begin{pmatrix}\phi&b_-\\b_+&c\end{pmatrix}\,,\qquad
 \psi=\begin{pmatrix}\psi^-\\\rho\end{pmatrix}\,,\qquad
 \psi'=\begin{pmatrix}\rho'\\\psi^+\end{pmatrix}\,,\qquad
 \cF=\begin{pmatrix}0&-\partial_-\phi\\
             \partial_+\phi&\partial_+b_-+\partial_-b_+\end{pmatrix}\,.
 \label{SMcandidatecomponents}
\ee
Then the action and propagating equations become
\be
 \boxed{\cL_\partial
   =\partial_+\phi\,\partial_-\phi
      +\psi^+\partial_-\psi^+
      +\psi^-\partial_+\psi^-\,,\qquad
 \partial_+\partial_-\phi=0\,,\quad
 \partial_-\psi^+=0\,,\quad\partial_+\psi^-=0\,.}
 \label{SMcandidatecomponentaction}
\ee
The propagating sector contains one nonchiral boson and two chiral
fermions. On this fixed background, the remaining components
$c,b_\pm,\rho,\rho'$ do not enter the action and admit arbitrary shifts.
Their correlators are not fixed by~\eqref{SMcandidatecomponentaction};
retaining them in quantum background responses requires an additional prescription.

On the propagating fields, the Kosmann transformations reduce to
\be
 \delta_v\phi=v^\mu\partial_\mu\phi\,,\qquad
 \delta_v\psi^\pm=v^\mu\partial_\mu\psi^\pm
                  +\half(\partial_\pm v^\pm)\psi^\pm\,.
 \label{SMcandidatecomponentWitt}
\ee
The RR gauge symmetry acts trivially on these fields:
\be
 \delta_\lambda\phi=0\,,\qquad \delta_\lambda\psi^\pm=0\,.
 \label{SMcandidatecomponentRR}
\ee

The solutions of~\eqref{SMcandidateparameters} and their weight-zero
Kosmann transformations are
\be
 \varepsilon=\begin{pmatrix}\epsilon^+(x^+)\\0\end{pmatrix}\,,\qquad
 \varepsilon'=\begin{pmatrix}0\\\epsilon^-(x^-)\end{pmatrix}\,,\qquad
 \delta_v\epsilon^\pm
 =v^\pm\partial_\pm\epsilon^\pm-\half(\partial_\pm v^\pm)\epsilon^\pm\,.
 \label{SMcandidatecomponentparameters}
\ee
These arbitrary chiral solutions are independent of $d$.
The transformations~\eqref{SMcandidatefermionic} become
\be
 \delta_\epsilon\phi=\psi^+\epsilon^++\psi^-\epsilon^-\,,\qquad
 \delta_\epsilon\psi^\pm=\epsilon^\pm\partial_\pm\phi\,.
 \label{SMcandidatecomponentSUSY}
\ee
Its off-shell variation is
\be
 \delta_\epsilon\cL_\partial
 =\partial_+(\psi^+\epsilon^+\partial_-\phi)
  +\partial_-(\psi^-\epsilon^-\partial_+\phi)\,,\qquad
 v^\pm=2\epsilon_1^\pm\epsilon_2^\pm
 \quad\hbox{in~\eqref{SMcandidateclosure}}\,.
 \label{SMcandidatecomponentSUSYL}
\ee
The supersupersymmetry algebra closes on the propagating fields as
\be
 \begin{gathered}
 \relax[\delta_1,\delta_2]\phi=v^\mu\partial_\mu\phi\,,\\
 [\delta_1,\delta_2]\psi^\pm\simeq
 v^\mu\partial_\mu\psi^\pm+\half(\partial_\pm v^\pm)\psi^\pm\,,
 \end{gathered}
 \label{SMcandidatecomponentclosure}
\ee
where $\simeq$ uses the fermion equations of motion.
For arbitrary even $\alpha_0,\alpha_\pm$, trivial fermionic symmetries are
\be
 \delta\psi^+=\alpha_0\partial_+\psi^-+\alpha_+\partial_-\psi^+\,,
 \qquad
 \delta\psi^-=\alpha_0\partial_-\psi^++\alpha_-\partial_+\psi^-\,,
 \qquad\delta\phi=0\,.
 \label{SMcandidatetrivial}
\ee
Their Lagrangian variation is
$\partial_-(\psi^+\delta\psi^+)+\partial_+(\psi^-\delta\psi^-)$,
and they vanish on shell.

The propagating sector of~\eqref{SMcandidatecomponentaction} is a free
${\mathcal N=(1,1)}$ scalar multiplet, providing a building block for
symmetric-product CFTs. Four copies, with the four bosons compactified
on $T^4$, yield the ${\mathcal N=(4,4)}$ torus seed of
$\operatorname{Sym}^N(T^4)$~\cite{Eberhardt:2018ouy,Eberhardt:2019ymp}.
A holographic identification of this classical
boundary CFT remains open.
\endgroup
\end{widetext}

\end{document}